\documentclass[preprints,article,accept,oneauthor,pdftex]{mdpi}
\firstpage{1} 
\pubvolume{xx}
\issuenum{1}
\articlenumber{5}
\pubyear{2019}
\copyrightyear{2019}
\history{Received: date; Accepted: date; Published: date}

\pdfoutput=1

\Title{Thermodynamic Neural Network}

\Author{Todd Hylton $^{1,\dagger}$\orcidA{}}

\AuthorNames{Todd Hylton}

\address{%
$^{1}$ \quad University of California, San Diego; thylton@ucsd.edu\\}

\corres{Correspondence: thylton@ucsd.edu}

\firstnote{Current address: 9500 Gilman Dr., La Jolla CA, 92093}

\abstract{A thermodynamically motivated neural network model is described that self-organizes to transport charge associated with internal and external potentials while in contact with a thermal reservoir. The model integrates techniques for rapid, large-scale, reversible, conservative equilibration of node states and slow, small-scale, irreversible, dissipative adaptation of the edge states as a means to create multiscale order.  All interactions in the network are local and the network structures can be generic and recurrent.  Isolated networks show multiscale dynamics, and externally driven networks evolve to efficiently connect external positive and negative potentials. The model integrates concepts of conservation, potentiation, fluctuation, dissipation, adaptation, equilibration and causation to illustrate the thermodynamic evolution of organization in open systems.  A key conclusion of the work is that the transport and dissipation of conserved physical quantities drives the self-organization of open thermodynamic systems.}


\begin{document}

\section{Introduction} \label{sec:introduction}
Applying concepts from thermodynamics and statistical physics to neural network models has a relatively long history, much of it centered on models of interacting spins on a lattice – Ising models.  The dynamics of these systems near critical points \cite{glauber1963time}\cite{suzuki1968dynamics} have been studied to derive, for example, magnetic response functions using mean field approximations.  The statistical properties of randomly disordered Ising models - spin glasses – are also well studied and understood in limiting cases including infinite range interactions \cite{kirkpatrick1978infinite}.  These ideas were extended to networks capable of storing memories as attracting states recalled using only a portion of the initial memory – Hopfield Networks \cite{hopfield1982neural}.  The statistical mechanics of these networks have been developed in detail, particularly regarding their capacity to store and recall memories \cite{amit1985spin}\cite{bruce1987dynamics}\cite{sompolinsky1988statistical}\cite{gutfreund1990statistical}.  Further development of these ideas to include “hidden” spin states not directly determined by a set of training vectors and thermodynamically inspired techniques for training these networks were captured in a class of models called Boltzmann Machines \cite{ackley1985learning}\cite{hinton1986learning}.  Statistical physics also found application in non-Ising neural network models including layered networks \cite{levin1990statistical}, complex network model spaces \cite{albert2002statistical}\cite{watkin1993statistical}, and directed, Markovian neuronal networks \cite{clark1988statistical}.  The motivations and techniques employed in this work borrow heavily from this history and the work on Ising models in particular.    

As compared to these earlier works, the Thermodynamic Neural Network (TNN) model presented here is distinguished by its electric circuit inspiration and the incorporation of physical concepts such as charge conservation, potential diffusion, reaction kinetics and dissipation-driven adaptation.  In particular, the transport of a conserved quantity does not appear as a primary concern in these earlier works.  Although learning is the primary objective of the model, the motivation is not the statistical generalization of a training set, the replication of a function, or the storage of a memory; rather, learning is viewed as adaptation to improve equilibration with external potentials and a thermal reservoir.  

The larger inspiration for this work is the long standing hypothesis that evolution of the natural world, including life, is driven by thermodynamics and constrained by the laws of physics \cite{schrodinger1944physical}\cite{schneider1994life}.  In particular the model addresses the case of external boundary potentials that vary slowly as compared to the characteristic equilibration time of the network and the resulting evolution of the network to minimize internal entropy production \cite{glansdorff1964general}, while creating entropy at the boundaries via the transport of conserved quantities \cite{schlogl1967statistical}.  These are the essential physics that drive the self-organization of the network.  A particular contribution of this work is the recognition that selective internal dissipation of conserved quantities is the means by which the system “learns” without “forgetting” earlier configurations that were effective under other boundary potentials.  Initially this work was inspired by experiments on collections of metallic balls in oil that self-assemble to create electrical connections when subject to external potentials \cite{junformation}.  Although the primary context for the work is physics and thermodynamics, we were also inspired by ideas from complex systems, chemistry, neurobiology, computation and cognitive science.
 
\section{Results} \label{sec:results}
The model comprises a collection of nodes connected by symmetrically weighted edges in contact with a thermal reservoir and driven by a collection of external biasing nodes through which external potentials may be applied.  Sec.\ref{sec:model concepts} presents a qualitative description of the concepts that underpin the model, which in Sec.\ref{sec:model details} are described mathematically.  Computer simulations implementing the model are described in Sec.\ref{sec:model simulations}.

\subsection{Model Concepts} \label{sec:model concepts}

\begin{figure}[H]
\centering
\includegraphics[width=0.9\columnwidth]{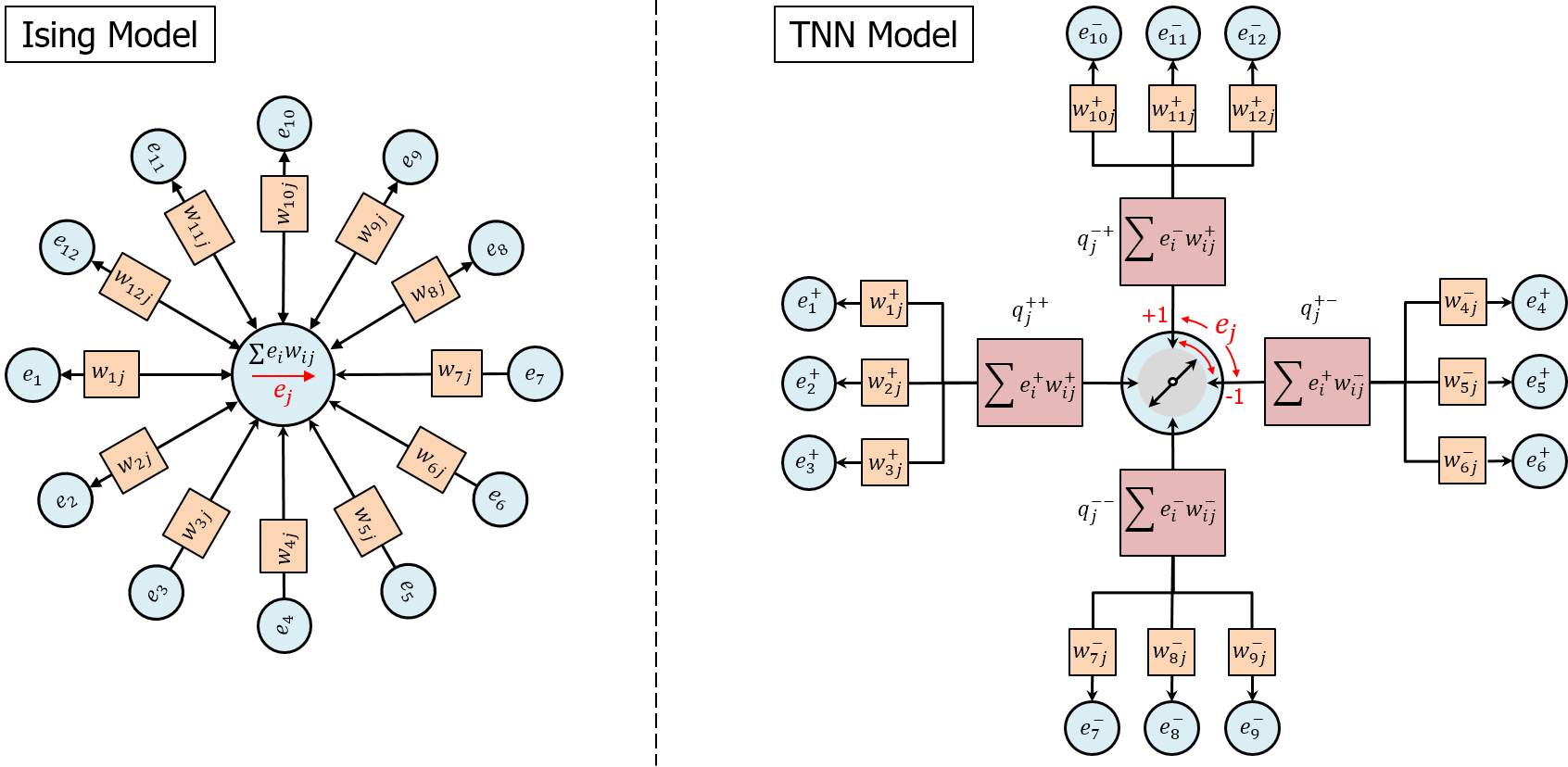}
\caption{(\textbf{left}) The Ising model is a network of nodes in which the node states ("spins") \(e_i\) interact via symmetric weights \(w_{ij}\).  Low energy states are those that align the node state \(e_j\) with the net total of the weighted interactions.  (\textbf{right}) The TNN model is also a network of nodes in which the node states ("potentials") \(e_i\) interact via symmetric weights \(w_{ij}\), but nodes interact via exchange of charge and low energy node states are those that effectively transport charge among the node’s inputs while independently conserving positive and negative charge.  Node state selection is a competition to connect two different pairs of compartments having opposite charge polarity and the same node input state polarity, illustrated as rotating switch.}
\label{fig:TNNvsIsing}
\end{figure}

Referring to Fig.\ref{fig:TNNvsIsing}, we compare the TNN model with the well known Ising model. The Ising model is a network of nodes in which the node states \(e_i\), typically conceived as "spins", interact via symmetric weights \(w_{ij}\).  Low energy states are those that align the node state \(e_j\) with the net total of the weighted interactions with its connected nodes, which can be viewed as accumulating all of the interactions \(\sum_i e_iw_{ij}\) into a single compartment. The TNN model is also a network of nodes in which the node states \(e_i\), conceived as electrical potentials, interact via symmetric weights \(w_{ij}\), but nodes interact via exchange of charge and low energy node states are those that effectively transport charge among the node’s inputs while independently conserving both positive and negative charges. 

Referring now to Fig.\ref{fig:TNNvsIsing} (\textbf{right}), the following list describes the concepts whereby the TNN model communicates and updates network node states.
\begin{itemize} \setlength{\parskip}{1em}
  \item A node \(j\) is characterized by a state \(e_j\) representing a \emph{potential}.  In general, the model supports any number of values of the node state on the interval \(e_j\in [\, -1, 1]\,\). For example, a binary node might assume a state \(e_j\in \{ -1, 1\}\). 
  \item An edge connecting nodes \(i\) and \(j\) is characterized by a real, symmetric \emph{weight} \(w_{ij}\) describing a capacity to transport \emph{charge}.
  \item Node \(i\) may apply a potential \(e_i\) to an edge weight \(w_{ij}\) and generate an edge charge \(q_{ij}=e_iw_{ij}\) that becomes input to node \(j\).  Similarly, node \(j\) may apply a potential \(e_j\) to the same edge weight and generate an edge charge  \(q_{ji}=e_jw_{ij}\) that becomes input to node \(i\).  In order to clarify the relationship between potentials and edges, we may sometimes designate potentials with 2 subscripts as \(e_{ij}=e_i\) and \(e_{ji}=e_j\).
  \item Positive and negative charges are independently conserved (i.e. they never sum to cancel each other) and communicate potential along which charges of opposite polarity should flow. By this means, externally applied potentials are able to diffuse through the network and connect to complementary potentials.  
  \item For a given node \(j\), charge conservation requires the aggregation of input node charges into 4 compartments, \(q_j^{\pm\pm} =\sum_i e_i^\pm w_{ij}^\pm\), distinguished by the signs of potentials \(e_i^\pm\) and weights \(w_{ij}^\pm\) that create the charge.  Depending on the inputs, anywhere from 1 to 4 compartments may be populated with charge at the time of the node state decision.
  \item Node state selection optimizes the transfer of charge between pairs of competing compartments using Boltzmann statistics, illustrated as a "switch" in Fig.\ref{fig:TNNvsIsing}.
  \item If two nodes are connected by an edge, then configurations in which the nodes have opposite potentials will be favored.  Hence, if the nodes are arranged on a regular grid and connected locally, then domains of "anti-ferromagnetic" order typically emerge.  
\end{itemize}

\begin{figure}[H]
\centering
\includegraphics[width=0.9\columnwidth]{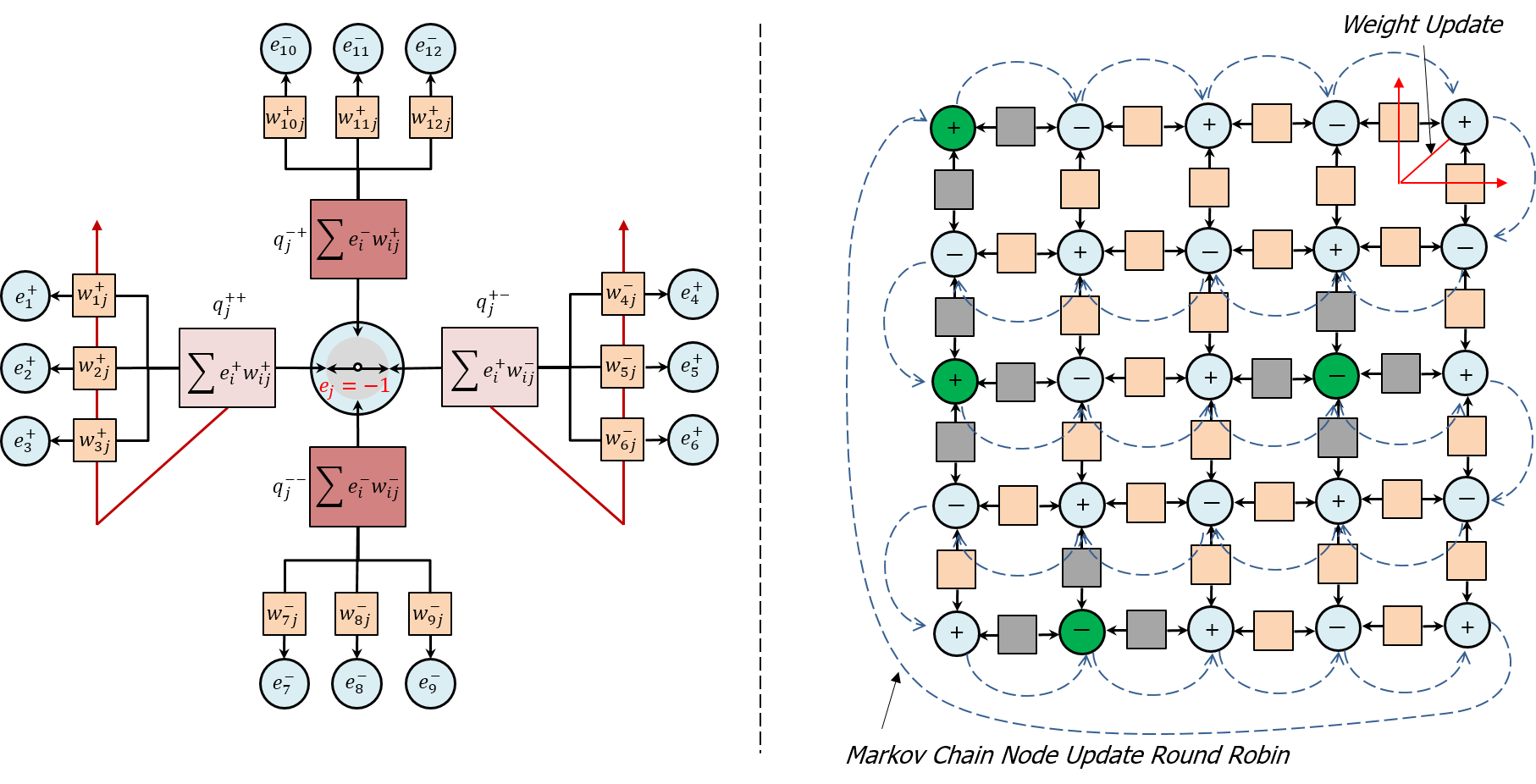}
\caption{(\textbf{left}) Node state selection results in a "setting of the switch" that connects a pair of \emph{selected} compartments (light pink) and relieves accumulated charge while \emph{unselected} compartments (dark pink) are unchanged. Edge states associated with the selected compartments adapt through \emph{dissipation} of residual compartment charge (red arrows). (\textbf{right}) A network of internal, adapting nodes (light blue) and external, biased nodes (green). A round robin Markov chain method (dashed blue arrows) continuously updates node states, while edge weights update only when a node is near equilibrium (red arrows). In order to transport complementary charge, node states favor connections to nodes with opposite polarities (nearest neighbor network shown).}
\label{fig:TNN_update}
\end{figure}

Referring now to Fig.\ref{fig:TNN_update} (\textbf{left}), node state selection results in a "setting of the switch" that connects a pair of \emph{selected} compartments and relieves accumulated charge by transporting complementary charges through the compartments as output to their connected nodes.  In this way, positive and negative charges are conserved and complementary conduction pathways are created through the network. The following list describes these ideas in more detail.
\begin{itemize} \setlength{\parskip}{1em}
  \item The node state decision selects and connects two compartments and transfers complementary charge between them, while leaving unselected compartments disconnected. If the compartments are populated such that no complementary pairs exist, then the state selection becomes uniformly random.  \emph{Residual charge} is that portion of the charge remaining on the selected compartments after charge transfer, which is in general unavoidable owing to the thermal fluctuations in the network.
  \item When a node is near equilibrium, residual charge in the selected compartments is dissipated to the thermal reservoir via updates to the associated edge weights according to a Boltzmann distribution.  In general, these updates improve the ability of the node to transport charge among the selected compartments if similar conditions are encountered in the future.
  \item Edge weights associated with unselected compartments are not updated and charge accumulated in the unselected compartments is retained.  In this way the node retains memory of "contrary" inputs (charges) and correlations (weights) that may be important to future decisions, which is intended to address the long standing "forgetting problem" in artificial neural networks.  
\end{itemize}

Referring now to Fig.\ref{fig:TNN_update} (\textbf{right}), A network of internal, adapting nodes and external, biased nodes that inject charge into the network is updated via a round robin Markov chain method that continuously updates node states, while edge weights update only when a node is near equilibrium. The following list describes these ideas in more detail.
\begin{itemize} \setlength{\parskip}{1em}
  \item Node states are updated in a continuous round-robin Markov chain, which guarantees that at the time of updating any particular node all the other nodes in the network have already updated and thereby preserves the idea of causality within the network interactions.
  \item Updates may be either reversible or irreversible, depending on the node's state of equilibration at the time of the update.  A node can determine its state of equilibration by examining the fluctuations in its energy over time.  If those fluctuations are small compared to the temperature, then the node may be deemed to be equilibrated and vice-versa.
  \item A reversible update, which happens when the node is non-equilibrated, corresponds to the node temporarily updating its compartments with new input charges and communicating its state to its connected nodes without updating edge weights.  The purpose of the reversible update is to generate fluctuations in the network that explore its configuration space to search for an equilibrium without destroying its previously acquired structure.
  \item An irreversible update, which happens when the node is equilibrated, corresponds to the node permanently updating its compartments with new input charges, updating the edge weights as described above, and communicating its state to its connected nodes.  The purpose of the irreversible update is to adapt the network to make it more effective at transporting charge in the future.
  \item This method of updating the network creates a continuous cycle of fluctuation, equilibration, dissipation and adaptation that connects and refines features at large spatial / short temporal scale (i.e. the collection of network node states), intermediate spatial / intermediate temporal scale (i.e. compartment charges) and small spatial / long temporal scale (i.e. edge weights) as the network evolves. In this way, the network can rapidly equilibrate to large scale changes in its environment through continuous refinement of its multiscale, internal organization.
  \item A range of network topologies is possible including multi-dimensional grid networks with near-neighbor connectivity, probabilistically connected, gridded networks with a metric that determines the probability of connection, and random networks.  In general, there is no imposition of hierarchy or “layers” upon the network as is common in most neural network models, but these kinds of networks can also be supported. Because connected nodes are driven to orient anti-ferromagnetically, most network configurations are inherently “frustrated” in that the nodes cannot find a way to satisfy this orientation with all their connected nodes.  For a special class of networks that are partitioned into two groups (bi-partitioned networks) in which nodes of one partition can connect only to nodes in the opposite partition, this frustration can be avoided.  Nearest neighbor grid networks are inherently bi-partitioned and are also attractive to study because they are easy to visualize.
\end{itemize}

As will be discussed in Sec.\ref{sec:model simulations}, isolated networks exposed only to a thermal bath can spontaneously order in ways reminiscent of solids, liquids and gases, while networks with externally biased nodes can self-organize in order to efficiently transport charge through the network.

\subsection{Model Details} \label{sec:model details}

\subsubsection{Network Model} \label{sec:network model}
The energy of network \(H_N\) is the sum of the node energies
\begin{equation} \label{eqn:network energy}
    H_N(\textbf{\textit{e}},\textbf{\textit{w}}, \textbf{\textit{q}}) = \sum_{j=1}^n H_j(e_j, \textbf{\textit{w}}_j, \textbf{\textit{q}}_j),
\end{equation}
where 
\renewcommand{\labelitemi}{-}
\begin{itemize} 
    \item \textbf{\textit{e}} refers to the set of node potentials \{\(e_1, e_2...e_n\)\},
    \item \textbf{\textit{w}} refers to the set of edge weights \{\(\textbf{\textit{w}}_1, \textbf{\textit{w}}_2...\textbf{\textit{w}}_n\)\}, 
    \item \(\textbf{\textit{w}}_j\) refers to the set of weights \{\(w_{1j}, w_{2j}...w_{m_jj}\)\} associated with node \textit{j},
    \item \textbf{\textit{q}} refers to the set of edge charges \{\(\textbf{\textit{q}}_1, \textbf{\textit{q}}_2...\textbf{\textit{q}}_n\)\}, 
    \item \(\textbf{\textit{q}}_j\) refers to the set of edge charges \{\(q_{1j}, q_{2j}...q_{m_jj}\)\} associated with node \textit{j},
    \item \(n\) refers to the number of nodes in the network, and
    \item \(m_j\) refers to the number of edges associated with node \(j\).
\end{itemize}
The network is assumed to be in contact with a thermal reservoir of inverse temperature \(\beta\), and the probability of a network state is assumed to follow Boltzmann statistics.
\begin{equation} \label{eqn:network equilibrium}
    P_N(\textbf{\textit{e}},\textbf{\textit{w}}, \textbf{\textit{q}}) = \frac{\exp{(-\beta H_N(\textbf{\textit{e}},\textbf{\textit{w}}, \textbf{\textit{q}}))}}{\sum_{\textbf{\textit{ewq}}}\exp{(-\beta H_N(\textbf{\textit{e}},\textbf{\textit{w}}, \textbf{\textit{q}}))}}
\end{equation}
In the following subsections, we develop a method to evolve the network toward the equilibrium of Eqn.\ref{eqn:network equilibrium} (even while the network is exposed to time varying external potentials) via a combination of reversible and irreversible updates as previously described. Reversible updates do not modify the edge states \textbf{\textit{w}} and \textbf{\textit{q}} and, thereby, sample the space of fluctuations in \textbf{\textit{e}} without modifying Eqn.\ref{eqn:network equilibrium}.  Conversely, irreversible updates do modify the edge states \textbf{\textit{w}} and \textbf{\textit{q}} and, thereby, also modify Eqn.\ref{eqn:network equilibrium}.

\subsubsection{Compartment Model} \label{sec:compartment model}
The charge conservation, complementary conduction pathways and state selection concepts previously described require the segregation of edge charges and weights according to the polarities of the potentials and weights that generate the charge.  We therefore segregate the edge charges as
\begin{equation} \label{eqn:edge charge}
    q_{ij}^{\pm\pm}=e_{ij}^\pm w_{ij}^{\pm\pm},
\end{equation}
where the superscripts refer to the polarity of the potential and weights associated with edge \(ij\).  Thus, for example, \(e_i^+\) refers to a positive input potential from node \(i\), \(q_{ij}^{+-}\) refers to an edge charge associated with a positive input potential from node \(i\) and a negative weight connecting nodes \(ij\), and  \(w_{ij}^{-+}\) refers to an edge weight associated with a negative input potential from node \(i\) and a positive weight connecting nodes \(ij\).  In the convention used here, where there are 2 polarity superscripts, the first refers to the sign of the input potential and the second to the sign of the edge weight. Edge charges are allowed to accumulate over multiple time steps, implying that a single edge may have more than one polarity segregated charge and weight at a given time.  The total charge that may accumulate on an edge is limited by the maximum node potentials \(e_i^\pm=\pm1\), which in Eqn.\ref{eqn:edge charge} may be realized as \(e_i^+=min\{\sum_te_i(t), 1\}\) and \(e_i^-=max\{\sum_te_i(t), -1\}\).

We further aggregate input charge, potential and weights into compartments by summing over the edges with common polarities of their input voltages and weights, thereby defining the following compartment input charges
\begin{equation*}
    q_j^{\pm\pm}=\sum_i q_{ij}^{\pm\pm}=\sum_{i \in \textbf{\textit{i}}^{\pm\pm}} q_{ij},
\end{equation*}
where \(\textbf{\textit{i}}^{\pm\pm}\) designates sets of edge indices with \(\pm\) input potentials and \(\pm\) edge weights, respectively.  Correspondingly, aggregating the output edge charges \(p_{ij}=e_j w_{ij}\) from node \(j\) after state selection \(e_j\), yields compartment output charges as
\begin{equation*}
    p_j^{\pm\pm}=e_jw_j^{\pm\pm}
\end{equation*}
where,
\begin{equation*}
    w_j^{\pm\pm}=\sum_i w_{ij}^{\pm\pm}=\sum_{i \in \textbf{\textit{i}}^{\pm\pm}} w_{ij}.
\end{equation*}

\subsubsection{Node Model} \label{sec:node model}
The choice of the node energy equation is driven by four considerations
\begin{itemize}
    \item minimizing residual charge on the node
	\item maximizing charge transport through the node
	\item avoiding attractors to node states \(e_j=0\)
	\item employing "kinetic" factors to sharpen state decisions and direct residual charge dissipation and accumulation processes
\end{itemize}
After much experimentation, the following expression for the node energy was selected
\begin{equation} \label{eqn:node energy 1}
\begin{split}
    H_j(e_j ) = &(f_j^+)^2 [(q_j^{++}-p_j^{+-})^2-(q_j^{++}+p_j^{+-})^2+(q_j^{+-}-p_j^{++})^2-(q_j^{+-}+p_j^{++})^2] \\ 
    + &(f_j^-)^2 [(q_j^{--}-p_j^{-+})^2-(q_j^{--}+p_j^{-+})^2+(q_j^{-+}-p_j^{--})^2-(q_j^{-+}+p_j^{--})^2]
\end{split}
\end{equation}
The first, third, fifth and seventh terms are associated with minimizing residual charge on the node.  The second, fourth, sixth and eighth terms are associated with maximizing charge transport.  The avoidance of attractors at \(e_j=0\) is addressed by the pairs of positive and negative terms that cancel terms in \((q_j^{\pm\pm} )^2\) and \((p_j^{\pm\pm} )^2\).  \(f_j^\pm=f_j^\pm(e_j)\) are the kinetic factors, which we describe later.  Expanding and collecting terms yields
\begin{equation} \label{eqn:node energy 2}
    \frac{H_j(e_j)}{4}=-(f_j^+)^2 (q_j^{++} p_j^{+-}+q_j^{+-} p_j^{++})-(f_j^-)^2 (q_j^{--} p_j^{-+}+q_j^{-+} p_j^{--})
\end{equation}
The terms in Eqn.\ref{eqn:node energy 2} are suggestive complementary pairs of “forces” \(q_j^{\pm\pm}\) and “fluxes”  \(p_j^{\pm\mp}\) that are familiar from thermodynamics.  We rewrite Eqn.\ref{eqn:node energy 2} as
\begin{equation} \label{eqn:node energy 3}
    \frac{H_j(e_j)}{4}=-(f_j^+)^2 (q_j^{++} w_j^{+-}+q_j^{+-} w_j^{++})e_j-(f_j^-)^2 (q_j^{--} w_j^{-+}+q_j^{-+} w_j^{--})e_j
\end{equation}
and recognize that each term in the node energy promotes the transfer of charge among compartments with opposite charge polarity and the node state selection is a competition between two pairs of compartments distinguished by the sign of their input potential.  The positive potential / positive weight / positive charge compartment, \(q_j^{++}\), and the positive potential / negative weight / negative charge component, \(q_j^{+-}\), form a pair that favors states \(e_j<0\), while negative potential / negative weight / positive charge compartment, \(q_j^{--}\), and the negative potential / positive weight / negative charge component, \(q_j^{-+}\), form a pair that favors states \(e_j>0\).   We note each term in Eqn.6 is the product of three factors: two factors that are determined by the compartment inputs, \(q_j^{\pm\pm}\) and \(w_j^{\pm\pm}\), and one that is determined by the node state, \(e_j\).  This enables the TNN to connect complementary potentials and update weights without reliance on carefully engineered network architectures and post hoc error assignment techniques like back-propagation, as will be explained below. 
\begin{figure} [H]
    \centering
    \includegraphics[width=0.7\columnwidth]{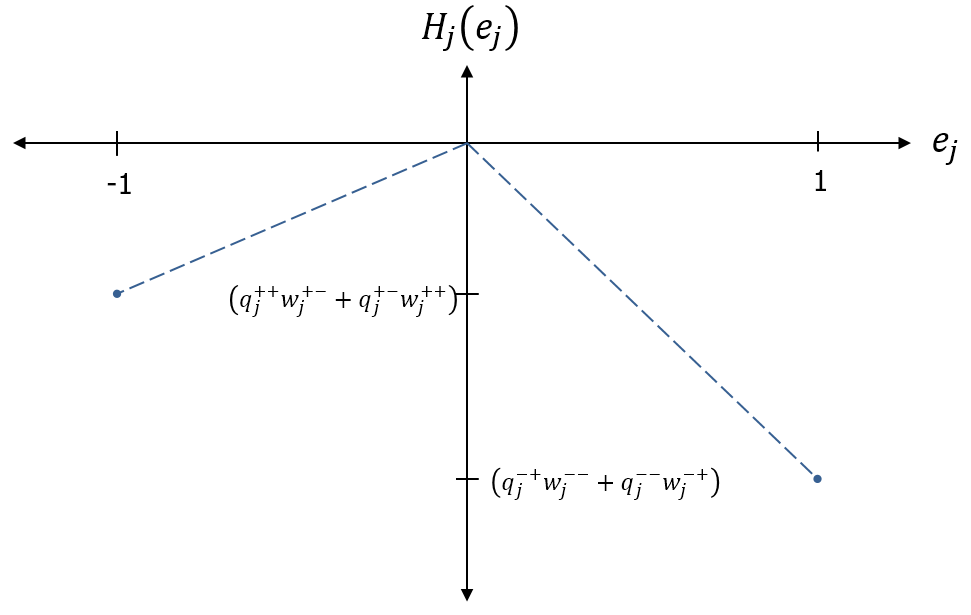}
    \caption{The energy landscape of a node using Heaviside kinetic factors, which promote state selection near \(e_j=\pm1\). The energy is piecewise linear in the state, which is a convenient choice that allows a simple spacing of node energies by a characteristic energy scale (temperature).}
    \label{fig:node energy landscape}
\end{figure}
The kinetic factors of Eqn.\ref{eqn:node energy 3} focus the updates to state variables in one of the compartment pairs.  The physical inspiration for the kinetic factors is that the state selection process, while “exciting” the reaction of the selected compartments also “inhibits” the reaction of the unselected of compartments.   A more colloquial interpretation is that the node can “address only one thing at a time” while “saving other things for later”.  We have chosen the name “kinetic” to connote the idea that this factor decides which compartments of charge should and should not “move” through the node.  We have experimented with different kinetic factors , but have found the following rectifying function to be effective
\begin{equation} \label{eqn:kinetic factors}
    f_j^\pm(e_j) = h(\mp e_j),
\end{equation}
where \(h\) is the Heaviside step function.  The energy landscape of the node using the kinetic factor of Eqn.\ref{eqn:kinetic factors} is shown in Fig.\ref{fig:node energy landscape}, illustrating how the kinetic factors sharpen the state selection.  As shown in the next section, kinetic factors also focus weight and charge updates on selected compartments, while protecting weight and charge associated with the unselected compartments.

\subsubsection{Edge Model} \label{sec:edge model}
For nodes executing an irreversible update, residual charge on the selected compartments is used to adapt the weights such that the node is more effective when encountering similar inputs in the future.  Selecting the residual charge terms from Eqn.\ref{eqn:node energy 1} we write the residual node energy after state selection as
\begin{equation} \label{eqn:residual charge energy}
\begin{split}
     H_j (\textbf{\textit{q}}_j,\textbf{\textit{w}}_j |e_j ) &= (f_j^+)^2 [(q_j^{++}-e_j w_j^{+-})^2+(q_j^{+-}-e_j w_j^{++})^2 ] \\ &+(f_j^-)^2 [(q_j^{--}-e_j w_j^{-+})^2+(q_j^{-+}-e_j w_j^{--})^2] \\ &= (-\delta_j^{++})^2+(-\delta_j^{+-})^2+(-\delta_j^{--})^2+(-\delta_j^{-+})^2   
\end{split}
\end{equation}
where
\begin{equation} \label{eqn:residual charge}
    -\delta_j^{\pm\pm}=f_j^\pm(q_j^{\pm\pm}-e_j w_j^{\pm\mp})
\end{equation}
are the residual compartment charges that we wish to minimize through weight updates \(w_{ij}\to w_{ij}+\Delta w_{ij}\).  We capture this objective by rewriting Eqn.\ref{eqn:residual charge energy} as
\begin{equation} \label{eqn:residual charge energy correction}
\begin{split}
        H_j (\Delta \textbf{\textit{w}}_j ) &= (\sum_{\textbf{\textit{i}}^{++}}e_i^+ \Delta w_{ij}^{++} - \sum_{\textbf{\textit{i}}^{+-}}e_j \Delta w_{ij}^{+-} - \delta_j^{++})^2 + (\sum_{\textbf{\textit{i}}^{+-}}e_i^+ \Delta w_{ij}^{+-} - \sum_{\textbf{\textit{i}}^{++}}e_j \Delta w_{ij}^{++} - \delta_j^{+-})^2 \\ &+ (\sum_{\textbf{\textit{i}}^{--}}e_i^- \Delta w_{ij}^{--} - \sum_{\textbf{\textit{i}}^{-+}}e_j \Delta w_{ij}^{-+} - \delta_j^{--})^2 + (\sum_{\textbf{\textit{i}}^{-+}}e_i^- \Delta w_{ij}^{-+} - \sum_{\textbf{\textit{i}}^{--}}e_j \Delta w_{ij}^{--} - \delta_j^{-+})^2
\end{split}
\end{equation}
As will be explained below, we elect to distribute the residual charges to the edges and rewrite Eqn.\ref{eqn:residual charge energy correction} as
\begin{equation} \label{eqn:residual charge split 1}
\begin{split}
    H_j (\Delta \textbf{\textit{w}}_j ) &= \left(\sum_{\textbf{\textit{i}}^{++}}(e_i^+ \Delta w_{ij}^{++} - \eta_{ij}^{++}) - \sum_{\textbf{\textit{i}}^{+-}}(e_j \Delta w_{ij}^{+-} - \mu_{ij}^{+-})\right)^2 \\ &+ \left(\sum_{\textbf{\textit{i}}^{+-}}(e_i^+ \Delta w_{ij}^{+-} - \eta_{ij}^{+-}) - \sum_{\textbf{\textit{i}}^{++}}(e_j \Delta w_{ij}^{++} - \mu_{ij}^{++})\right)^2 \\ &+ \left(\sum_{\textbf{\textit{i}}^{--}}(e_i^- \Delta w_{ij}^{--} - \eta_{ij}^{--}) - \sum_{\textbf{\textit{i}}^{-+}}(e_j \Delta w_{ij}^{-+} - \mu_{ij}^{-+})\right)^2 \\ &+ \left(\sum_{\textbf{\textit{i}}^{-+}}(e_i^- \Delta w_{ij}^{-+} - \eta_{ij}^{-+}) - \sum_{\textbf{\textit{i}}^{--}}(e_j \Delta w_{ij}^{--} - \mu_{ij}^{--})\right)^2
\end{split}
\end{equation}
where
\begin{equation} \label{eqn:charge split factors}
    \begin{split}
        &\eta_{ij}^{\pm\pm}= \frac{|e_i|}{\Delta_j^{\pm\pm}} \cdot \delta_j^{\pm\pm} \\ &\mu_{ij}^{\pm\mp}= \frac{|e_j|}{\Delta_j^{\pm\pm}} \cdot \delta_j^{\pm\pm} \\ &\Delta_j^{\pm\pm}=\sum_{\textbf{\textit{i}}^{\pm\pm}}|e_i| + \sum_{\textbf{\textit{i}}^{\pm\mp}}|e_j| 
    \end{split}
\end{equation}
The terms \(\Delta_j^{\pm\pm}\) represent the aggregated compartment input potentials and their paired output compartment potentials.  The residual charge distributions to the edges, \(\eta_{ij}^{\pm\pm}\) and \(\mu_{ij}^{\pm\mp}\), are the apportioned by their respective input potential's contribution to the \(\Delta_j^{\pm\pm}\).  In other words, the residual charge remaining on the selected compartments is distributed as "error" to their edge weights in proportion to the magnitude of their corresponding input potentials. This distribution of charge to the edge weights is the only way that the network can dispose of charge: the only way that it can "violate" charge conservation.  Hence, we refer to this process as a "dissipation" of residual charge to the edge the weights, which are adapted by it as we describe next.  

In the implementation described below, the weight updates are sampled independently such that the time average of the cross terms in Eqn.\ref{eqn:residual charge split 1} vanish.  Hence, we rewrite Eqn.\ref{eqn:residual charge split 1} as
\begin{equation} \label{eqn:residual charge split 2}
\begin{split}
    H_j (\Delta \textbf{\textit{w}}_j ) &\approx \sum_{\textbf{\textit{i}}^{++}}(e_i^+ \Delta w_{ij}^{++} - \eta_{ij}^{++})^2 + \sum_{\textbf{\textit{i}}^{+-}}(e_j \Delta w_{ij}^{+-} - \mu_{ij}^{+-})^2 \\ &+ \sum_{\textbf{\textit{i}}^{+-}}(e_i^+ \Delta w_{ij}^{+-} - \eta_{ij}^{+-})^2 + \sum_{\textbf{\textit{i}}^{++}}(e_j \Delta w_{ij}^{++} - \mu_{ij}^{++})^2 \\ &+ \sum_{\textbf{\textit{i}}^{--}}(e_i^- \Delta w_{ij}^{--} - \eta_{ij}^{--})^2 + \sum_{\textbf{\textit{i}}^{-+}}(e_j \Delta w_{ij}^{-+} - \mu_{ij}^{-+})^2 \\ &+ \sum_{\textbf{\textit{i}}^{-+}}(e_i^- \Delta w_{ij}^{-+} - \eta_{ij}^{-+})^2 + \sum_{\textbf{\textit{i}}^{--}}(e_j \Delta w_{ij}^{--} - \mu_{ij}^{--})^2
\end{split}
\end{equation}
Collecting terms, completing squares, dropping terms that are independent of \(\Delta w_{ij}\), and substituting the definitions from Eqn.\ref{eqn:charge split factors} yields
\begin{equation} \label{eqn:residual charge split 3}
\begin{split}
    H_j (\Delta \textbf{\textit{w}}_j ) &\approx \sum_{\textbf{\textit{i}}^{++}}(e_i^2+e_j^2)\left[\Delta w_{ij}^{++} - \frac{1}{e_i^2+e_j^2} \left(\frac{e_i}{|e_i|}\frac{\delta_j^{++}}{\Delta_j^{++}}e_i^2 - \frac{e_j}{|e_j|}\frac{\delta_j^{+-}}{\Delta_j^{+-}}e_j^2\right)\right]^2 \\ & + \sum_{\textbf{\textit{i}}^{+-}}(e_i^2+e_j^2)\left[\Delta w_{ij}^{+-} - \frac{1}{e_i^2+e_j^2} \left(\frac{e_i}{|e_i|}\frac{\delta_j^{+-}}{\Delta_j^{+-}}e_i^2 - \frac{e_j}{|e_j|}\frac{\delta_j^{++}}{\Delta_j^{++}}e_j^2\right)\right]^2 \\ & + \sum_{\textbf{\textit{i}}^{--}}(e_i^2+e_j^2)\left[\Delta w_{ij}^{--} - \frac{1}{e_i^2+e_j^2} \left(\frac{e_i}{|e_i|}\frac{\delta_j^{--}}{\Delta_j^{--}}e_i^2 - \frac{e_j}{|e_j|}\frac{\delta_j^{-+}}{\Delta_j^{-+}}e_j^2\right)\right]^2 \\ & + \sum_{\textbf{\textit{i}}^{-+}}(e_i^2+e_j^2)\left[\Delta w_{ij}^{-+} - \frac{1}{e_i^2+e_j^2} \left(\frac{e_i}{|e_i|}\frac{\delta_j^{-+}}{\Delta_j^{-+}}e_i^2 - \frac{e_j}{|e_j|}\frac{\delta_j^{--}}{\Delta_j^{--}}e_j^2\right)\right]^2
\end{split}
\end{equation}

The form of Eqn.\ref{eqn:residual charge split 3} is the motivation for the choice of distributing residual charge among the weights as in Eqn.\ref{eqn:charge split factors}.  The sum of quadratic terms, when exponentiated according to Boltzmann statistics, results in independent Gaussian distributions with simple offsets.  If the node states are constrained to be binary \(e_i=\pm1\), then the weight offsets are identical for every weight in an input compartment, which is perhaps the least biased of the possible choices for distributing the residual charge.  Because each weight is connected to two nodes, it receives an update as specified in Eqn.\ref{eqn:residual charge split 3} from each node.  In our implementation these two updates are separate sampling events, each providing \(\frac{1}{2}\) of the total update (i.e. the Gaussian offset is divided by two) in Eqn.\ref{eqn:residual charge split 3}. Also, though not detailed here, to lowest order these weight updates do not affect the charge transport terms in Eqn.\ref{eqn:node energy 1} that were ignored in Eqn.\ref{eqn:residual charge energy}.

The update of weights according the technique just described is convenient because the updates can be made independently, which greatly simplifies the computational model.  It does, however, introduce weight growth as an undesirable artifact.  If we consider the collection of weights \(\textbf{\textit{w}}_j\) associated with node \(j\) as a vector then the effect of this independent weight updating is to on average increase the magnitude, \(|\textbf{\textit{w}}_j|\), which can be seen from
\begin{equation*}
    \langle (\textbf{\textit{w}}_j + \Delta\textbf{\textit{w}}_j )^2 \rangle = \textbf{\textit{w}}_j^2 + \langle \Delta\textbf{\textit{w}}_j^2 \rangle =  \textbf{\textit{w}}_j^2 + \frac{n_j}{2\beta} >  \textbf{\textit{w}}_j^2,
\end{equation*}
where \(n_j\) is the number edges connected to node j that are being updated and \(\beta\) is the inverse temperature.  One solution to this problem that still allows the weight updates to be performed independently is to reduce the size of the weight update to account for this weight growth.  In our implementation, we reduce the size of each weight on every update by a factor
\begin{equation} \label{eqn:weight reduction}
    \textbf{\textit{w}}_j \to \textbf{\textit{w}}_j \left(\frac{1}{1+n_j/(2 \beta \textbf{\textit{w}}_j^2)}\right)^{1/2}
\end{equation}
Note that this correction becomes small as \(\textbf{\textit{w}}_j^2\) becomes large, but has large effect at small weight sizes. 

Because the residual charge terms in Eqn.\ref{eqn:residual charge} are proportional to the kinetic terms \(f_j^\pm\), in the most general case only a fraction of the residual charge will be dissipated as weight updates, in which case charge conservation requires that the undissipated fraction be retained as edge charge (which would influence future state decisions and weight updates).  Hence, as the edge weight updates are sampled according to Eqn.\ref{eqn:residual charge split 3}, in general the edge charges are updated as
\begin{equation}
    q_{ij}^{\pm\pm} \to (1-f_j^\pm)q_{ij}^{\pm\pm}
\end{equation}
In the case of the Heaviside kinetic factors considered here, this simplifies to 
\begin{equation*}
    q_{ij}^{\pm\pm} \to h(\pm e_j)q_{ij}^{\pm\pm},
\end{equation*}
which dissipates all the residual charge in the selected compartments to update their weights (while retaining all of the residual charge in the unselected compartments).

\subsubsection{Network Evolution Model} \label{sec:network evolution model}
The task of the network simulation is to evolve the network toward global low-energy states in the presence of time varying inputs and thermal fluctuations, which we implement with as a Markov chain in which each node state is sampled in a continuous round robin for the duration of the simulation.  Each cycle through the nodes is considered to be a “step” in the simulation (displayed as a frame in the videos of Sec.\ref{sec:model simulations}), although there is no discontinuity in the round robin update between steps. When the node j is reached in the round robin, its conditional state is sampled according to a Boltzmann distribution as
\begin{equation} \label{eqn:node state sampling}
\begin{split}
    P_j (e_j |\textbf{\textit{e}} \neq e_j,\textbf{\textit{w}},\textbf{\textit{q}}) = \frac{\exp{(-\beta H_j (e_j))}}{Z_j} \\ Z_j = \sum_{e_j} \exp{(-\beta H_j (e_j))} 
\end{split}
\end{equation}
After sampling a node state \(e_j\) according to Eqn.\ref{eqn:node state sampling}, the node makes a choice to update its connections in one of two ways.
\renewcommand{\labelitemi}{ }
\begin{itemize} \setlength{\parskip}{1em}
    \item \underline{Reversible update} - If the node’s energy is fluctuating between simulation steps by an amount larger than its temperature, then it has not achieved equilibrium with the network.  The node temporarily updates its compartment input charges, selects a state, and communicates its state to its connected nodes without updating its edge weights.  This update drives fluctuations in the network that feedback to the node as the surrounding network evolves to find a common, low energy configuration of node states.  Different network states are sampled in this process, but their energy distribution is unchanged, which makes this node update process “reversible”.  We refer to these updates as “global” or “large-scale” because their primary purpose is to drive the larger network to find a low energy configuration of node states. 
    \item \underline{Irreversible update} - If the node’s energy is fluctuating between simulation steps by an amount smaller than its temperature, then it has achieved equilibrium with the network.  The node permanently updates its input compartment charges, selects a state, communicates its state to its connected nodes, and dissipates residual charge in its selected compartments as updates to its edge weights (Eqn.\ref{eqn:residual charge split 3}).  This update permanently modifies the energy distribution of the network states, which makes it “irreversible”.  We refer to these updates as “local” or “small-scale” because their primary purpose is to improve the efficiency of each node by updating its edge weights.
\end{itemize}
The intuition behind this update strategy is that when fluctuations are high, the node should communicate with its connections reversibly until the surrounding network can settle into an equilibrium state.  After this settling, the node can “commit” to updates that will improve its performance in the future.   At the scale of the network, the idea is that a combination of rapid, global, reversible relaxation of the network node states combined with slower, local, irreversible relaxation of the edge states will connect large and small scale dynamics while using only local interactions to make the computations.  A familiar analogy is a meeting of people in which information is communicated until a common (low fluctuation) understanding is achieved, at which time individual commitments can be made to improve upon the current situation.

Additionally, the Markov chain round robin implementation of these reversible and irreversible updates (Fig.\ref{fig:TNN_update}) insures that the network dynamics obey causality, by (1) guaranteeing that the entire network is updated prior to any node update, and by (2) learning causal structure (as may be imposed by externally biased nodes) by updating edge states only near equilibrium.  Hence, the method addresses the long-standing issue of separating causation and correlation in statistics.  The underlying hypothesis employed here is that thermodynamic evolution always proceeds in the direction of (local) equilibrium and that causal structure in the external potentials becomes embodied within the resulting organization.

We note that network evolution in the TNN is continuous and online without separate passes for "learning" and "inference" found in most artificial neural network models.

\subsubsection{External Bias Model} \label{sec:external bias model}
Network models may also include external bias nodes that are sources and sinks of charge for the larger network.  In the results of Sec.\ref{sec:externally biased networks}, these bias nodes have predetermined node states (e.g. a temporal pattern of \(e_j=\pm1\)) and large, fixed output weights.  A single bias node is able to polarize proximally connected network nodes (i.e. create some order in the network in its vicinity) and, thereby, diffuse potential into the network.  When paired with another bias node of opposite polarity, these polarized regions can evolve to form a conducting bridge that connects them or a domain wall that separates them.

\subsubsection{Other Network Effects} \label{sec:network effects}
For most network topologies the nodes are “frustrated” because there is no way to achieve perfect antiferromagnetic order.  The result of this frustration is that networks segment themselves into "domains" in order to minimize energy and effectively transport charge.  Edges within domains adapt weights to remove residual charges and improve transport efficiency (these edges belong to selected compartments) while edges separating domains sustain charges and weights (these edges belong to unselected compartments).  Hence, the domain walls maintain both intermediate-term memory (as compartment charges) and long-term memory (as edge weights) of previous network configurations even as the network adapts to its current inputs.  These memories enable the network to rapidly adapt to inputs similar to those it has previously encountered.  The creation and destruction of domains in these frustrated networks is perhaps analogous the creation of “virtual networks” among collections of excitatory and inhibitory neurons in biological systems \cite{yufik2002mind}. In bi-partitioned networks, however, the geometric frustration just described can be largely avoided.  Nearest neighbor grid networks are inherently bi-partitioned because the nodes fall naturally into two groups on a “checkboard” pattern.  These networks are particularly attractive to study because they are easy to visualize.  As externally biased nodes are introduced into bi-partitioned networks, however, geometric frustration can once again emerge as the node biases may conflict depending on their placement in the partitions and their relative polarity.  These effects are elaborated in the Fig.\ref{fig:2 bias 900 node 4nn} below.


Different node and edge temperatures allow the exploration of temperature dependent ordering dynamics in unbiased networks.   Lower (higher) node temperature relative to edge temperature results in more (less) ordered networks.  Ordering in the unbiased networks is also very sensitive to the connectivity of the nodes – more connectivity typically yields greater order.  Externally biased nodes introduce other energy scales into the network that compete with thermal fluctuations to determine network dynamics and organization.

\subsection{Model Simulation} \label{sec:model simulations}
The following images are snapshots of the evolution of the network captured as images of the node states at the end of one complete cycle of node updates using the techniques described in Sec.\ref{sec:model details}.  These images are frames from videos that can be viewed by following the hyperlinks in the image description.  In the images that follow, each node is one square and its state is indicated on a grayscale with \textit{black}=-1 and \textit{white}=+1.  The examples below focus on 2-dimensional (bi-partitioned) networks with nearest neighbor (NN) connectivity and periodic boundary conditions (the left / right and top / bottom edges are connected) because these networks allow easy visualization of the network organization.  The ideas presented below also apply to higher dimensional networks and networks with more complex connectivity such as randomly connected networks.  In every case, the network is initialized with random node and edge state values and allowed to evolve according to the methodology of the previous section.

\subsubsection{Isolated Networks} \label{sec:isolated networks}
\begin{figure} [H]
    \centering
    \includegraphics[width=0.3\columnwidth]{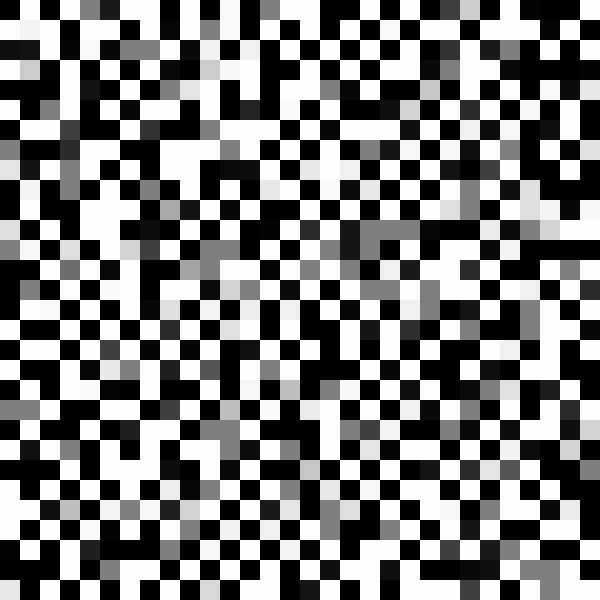}
    \includegraphics[width=0.3\columnwidth]{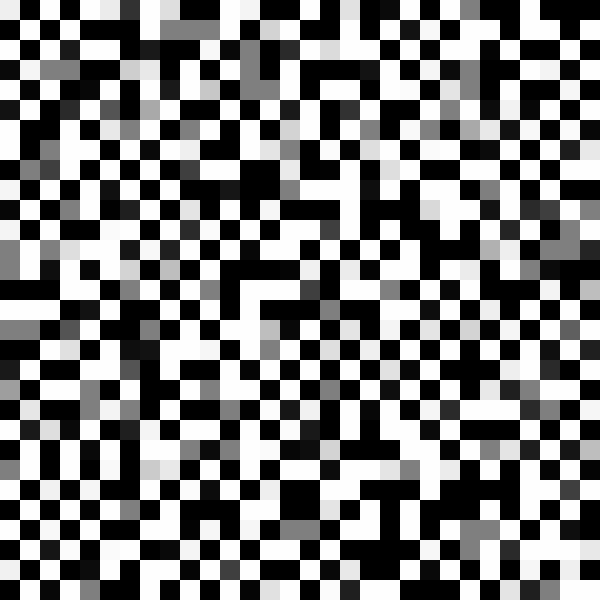}
    \includegraphics[width=0.3\columnwidth]{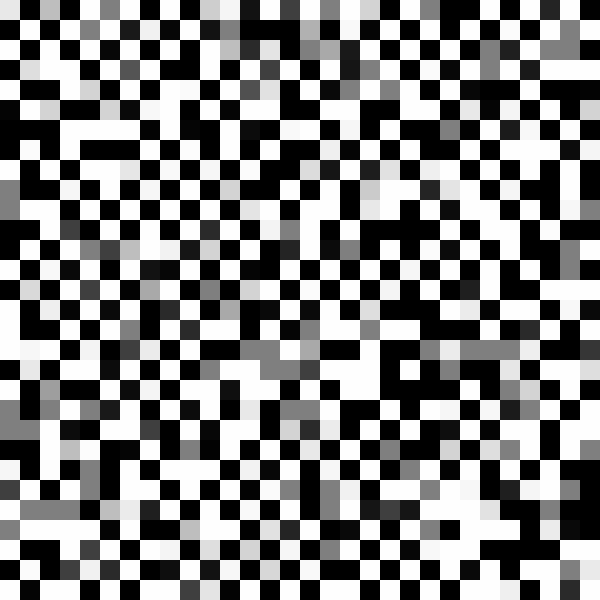}
    \caption{Snapshots from the evolution of a two dimensional, bi-partitioned network of 900 nodes with each node connected to its 4 nearest neighbors in the opposite partition with \(T_{node} / T_{edge} = 1\).  Networks are dynamic and noisy owing to contact with the thermal bath.  The propensity for the nodes to organize anti-symmetrically is evident in the checkboard appearance of the various regions of the images. Video S1 and at \url{https://youtu.be/3WFO41aV9lg}.}
    \label{fig:isolated w/ antiferro domains}
\end{figure}

\begin{figure} [H]
    \centering
    \includegraphics[width=0.3\columnwidth]{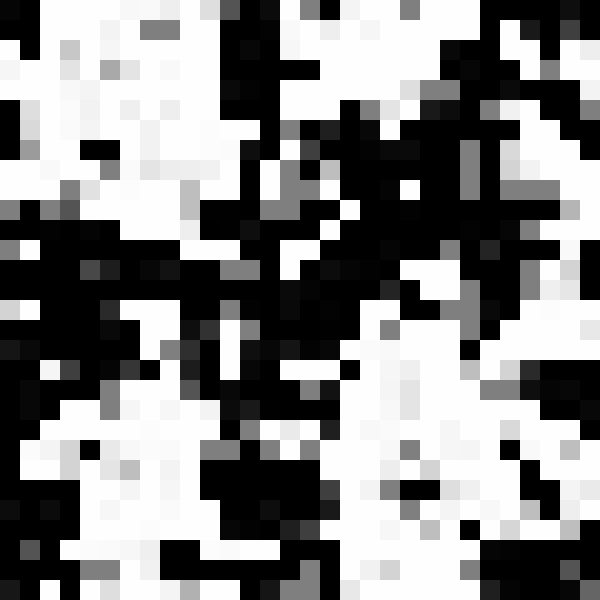}
    \includegraphics[width=0.3\columnwidth]{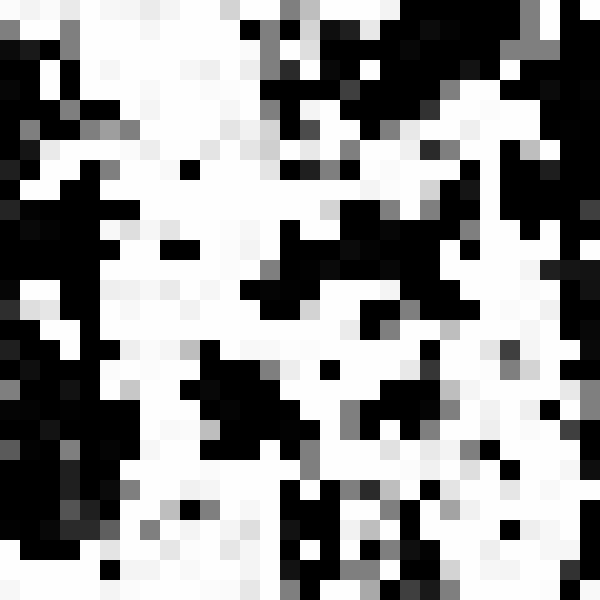}
    \includegraphics[width=0.3\columnwidth]{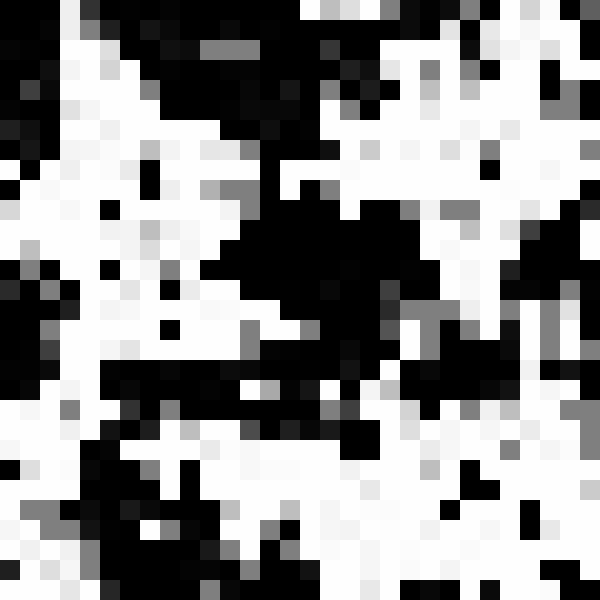}
    \caption{A different simulation of the same network as in Fig.\ref{fig:isolated w/ antiferro domains}. Because antisymmetric order is challenging to visualize, in these images the order is displayed by reversing the sign of the node state in one of the partitions (i.e. on every other square on the checkboard).  This change is applied only to the image display: the underlying order is still antiferromagnetic.  When displaying images this way domains appear as preferentially “white” or “black”. Video S2 and at \url{https://youtu.be/8_dvWLFr4mA}.}
    \label{fig:isolated w/ ferro domains}
\end{figure}

Figs.\ref{fig:isolated w/ antiferro domains} and \ref{fig:isolated w/ ferro domains} are sample results from unbiased networks with with 4 nearest neighbor (4NN) connectivity interacting with a thermal bath, illustrating the propensity of the network to organize. At the node and edge temperatures \(T_{node} / T_{edge} = 1\) selected for these simulations, ordering is local and transient. Fig.\ref{fig:isolated w/ ferro domains} inverts the display polarity of one the network partitions in order to show the ordering as “ferromagnetic” instead of the “anti-ferromagnetic” (this convention is adopted in all figures except Fig.\ref{fig:isolated w/ antiferro domains} for bi-partitioned networks).    


\begin{figure} [H]
    \centering
    \includegraphics[width=0.3\columnwidth]{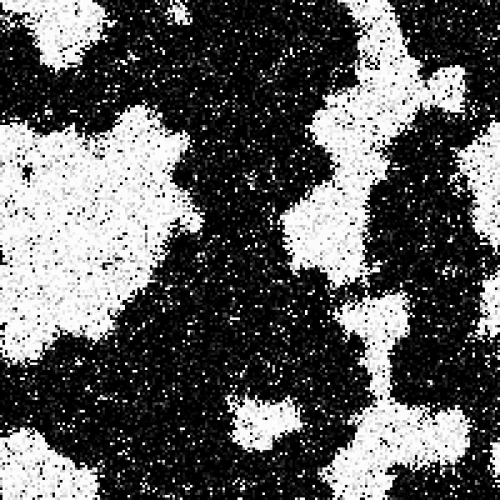}
    \includegraphics[width=0.3\columnwidth]{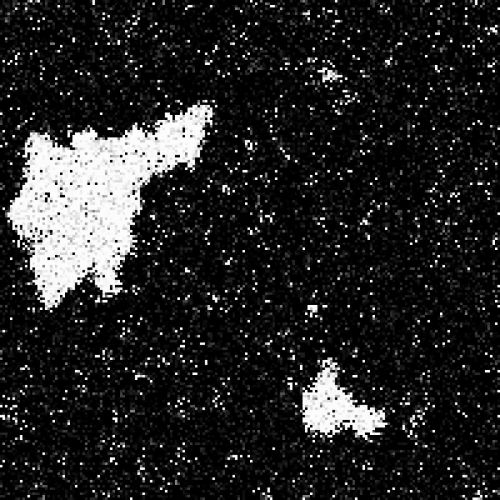}
    \includegraphics[width=0.3\columnwidth]{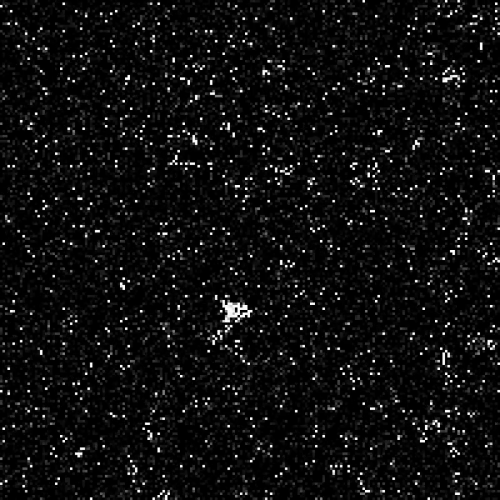}
    \caption{Three frames from the evolution (early to late from left to right) of a two dimensional, bi-partitioned network of 40,000 nodes with each node connected to its 16 nearest neighbors in the opposite partition at \(T_{node} / T_{edge} = 80\). The network evolves to a large single domain with local excitations. Video S3 and at \url{https://youtu.be/1Dj59El93KE}.}
    \label{fig:isolated 40k 16nn 80T}
\end{figure}

\begin{figure} [H]
    \centering
    \includegraphics[width=0.3\columnwidth]{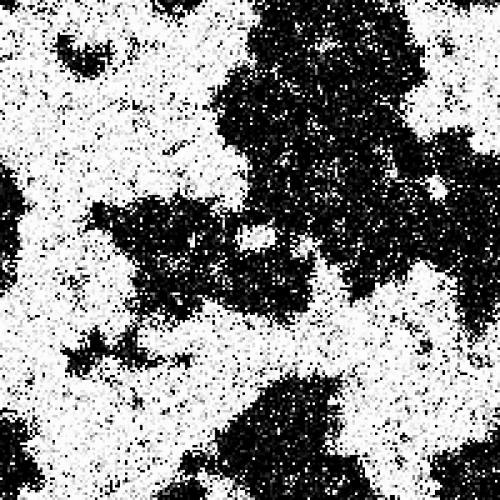}
    \includegraphics[width=0.3\columnwidth]{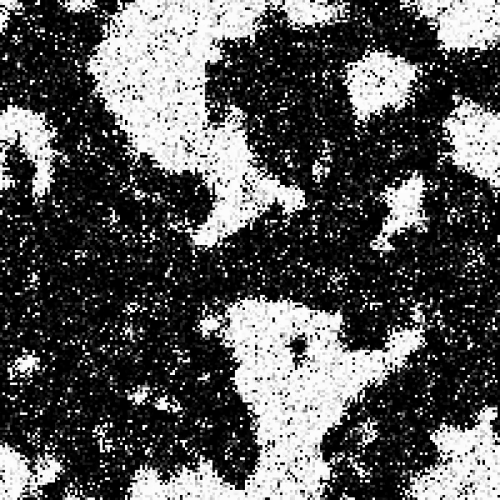}
    \includegraphics[width=0.3\columnwidth]{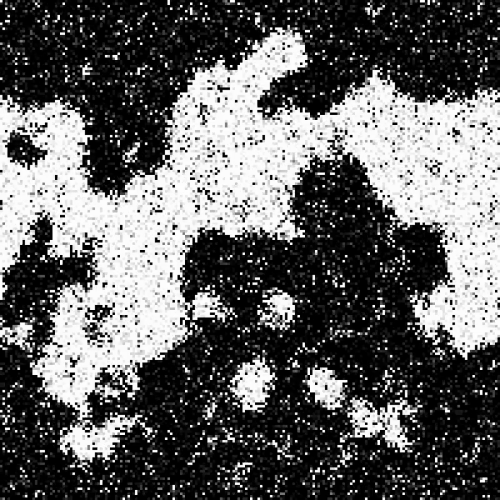}
    \caption{Three frames in the evolution of the same network as in Fig.\ref{fig:isolated 40k 16nn 80T} at \(T_{node} / T_{edge} = 100\). The network evolves complex multiscale dynamics with domain nucleation and growth. Video S4 and at \url{https://youtu.be/Ca9XEGyBytg}.}
    \label{fig:isolated 40k 16nn 100T}
\end{figure}

\begin{figure} [H]
    \centering
    \includegraphics[width=0.3\columnwidth]{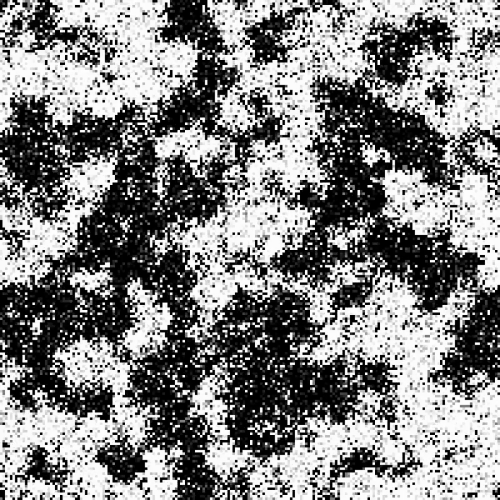}
    \includegraphics[width=0.3\columnwidth]{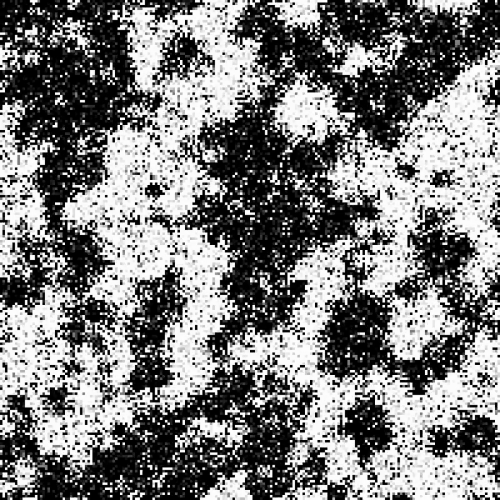}
    \includegraphics[width=0.3\columnwidth]{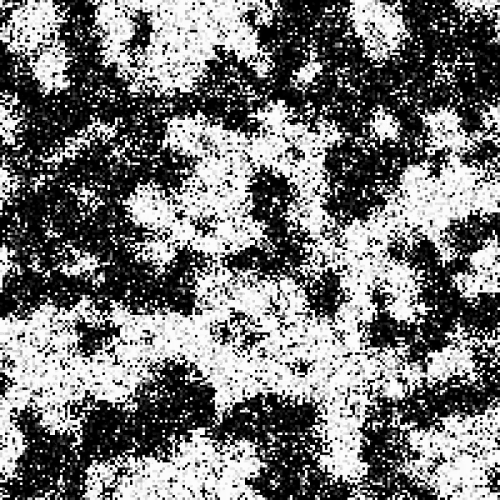}
    \caption{Three frames in the evolution of the same network as in Fig.\ref{fig:isolated 40k 16nn 80T} at \(T_{node} / T_{edge} = 120\). The network is dominated by faster, smaller scale dynamics as compared to Fig.\ref{fig:isolated 40k 16nn 100T}. Video S5 and at \url{https://youtu.be/UUA08xwcMAQ}.}
    \label{fig:isolated 40k 16nn 120T}
\end{figure}

\begin{figure} [H]
    \centering
    \includegraphics[width=0.4\columnwidth]{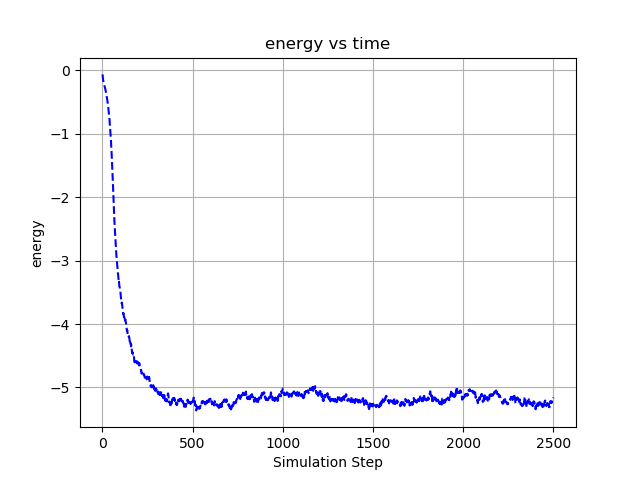}
    \includegraphics[width=0.4\columnwidth]{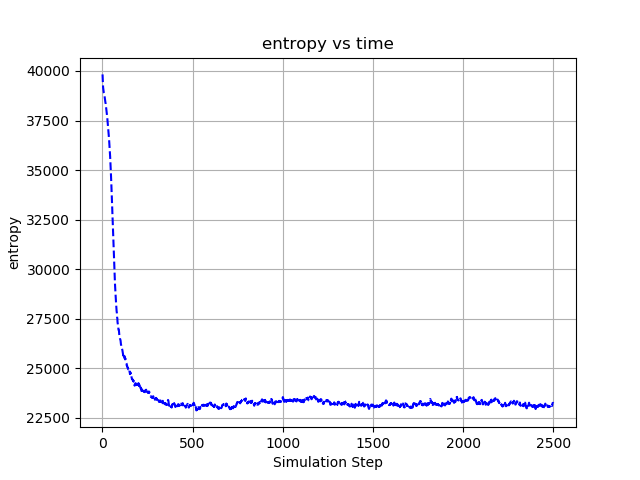} \par
    \includegraphics[width=0.4\columnwidth]{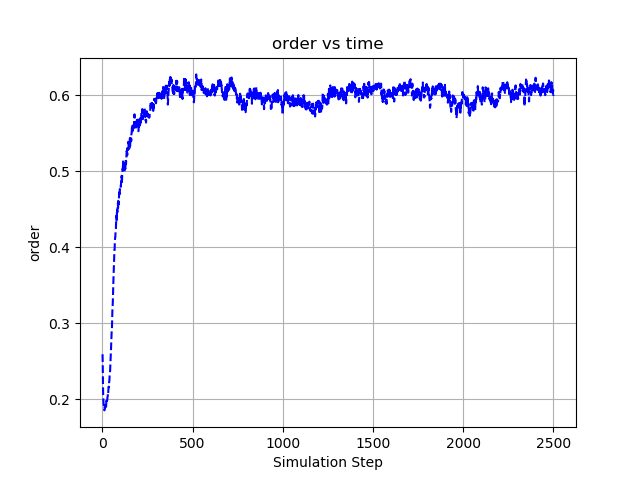}
    \includegraphics[width=0.4\columnwidth]{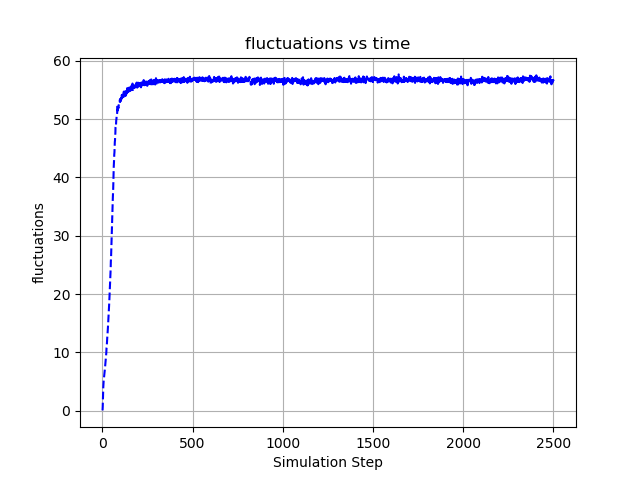}
    \caption{Temporal evolution of selected network averages for the simulation of Fig.\ref{fig:isolated 40k 16nn 100T} – node energy, node entropy, order, and fluctuations vs simulation time.  Energy is the node energy of Eqn.\ref{eqn:node energy 2} averaged over all the nodes.  Entropy is the sum of the nodes entropies derived from Eqn.\ref{eqn:node state sampling} normalized to a maximum value of 40,000.  Order is average over all edges of (the negative of) the product of the edge’s connected node states.  Fluctuations are the percentage of time that nodes choose a reversible update averaged over all the nodes.  Ordering in the network is consistent with decreased node energy and entropy and increased order and reversible node updates.}
    \label{fig:isolated 40k 16nn 100T statistics}
\end{figure}

Figs.\ref{fig:isolated 40k 16nn 80T}, \ref{fig:isolated 40k 16nn 100T}, and \ref{fig:isolated 40k 16nn 120T} shows the evolution of larger bi-partitioned networks with 16 nearest neighbor connections per node at 3 different node-to-edge temperatures chosen to illustrate the dynamics of the network with different levels of thermal excitation.  All simulations start from a highly disordered state (not shown in figures) and evolve to display complex, multiscale dynamics.  In general, as the node-to-edge temperature decreases, ordering extends over larger spatial and temporal scales.  Fig.\ref{fig:isolated 40k 16nn 100T statistics} shows plots of selected network statistics capturing the ordering of the network of Fig.\ref{fig:isolated 40k 16nn 100T}.


\begin{figure} [H]
    \centering
    \includegraphics[width=0.29\columnwidth]{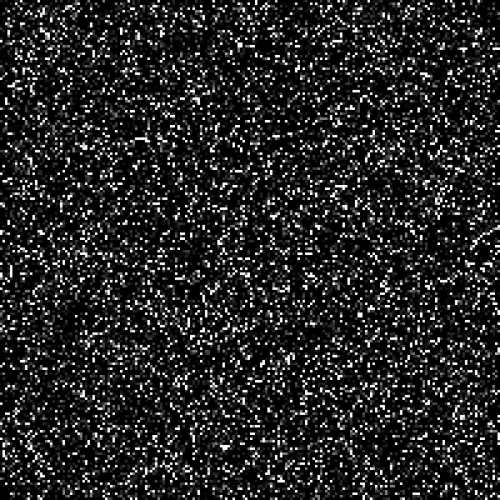}
    \includegraphics[width=0.29\columnwidth]{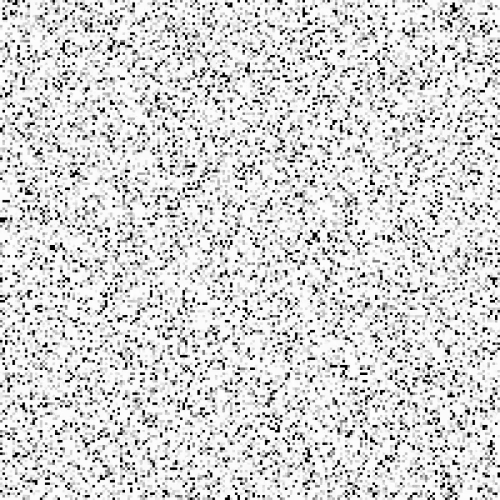}
    \includegraphics[width=0.4\columnwidth]{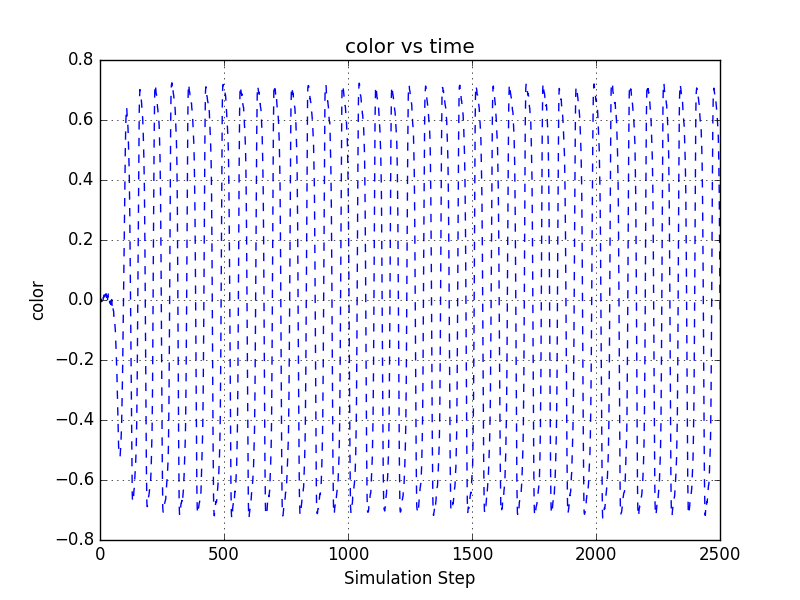}
    \caption{A bi-partitioned network of 40,000 nodes randomly connected to 16 nodes in the opposite partition at \(T_{node} / T_{edge} = 120\) uniformly oscillates periodically between mostly dark and light states. (\textbf{left and center}) Two frames from the evolution (\href{https://youtu.be/1YA5xauI5Y0}{video}) of the network showing dark and light states. (\textbf{right}) \textit{Color} shows the average state value of the network nodes and illustrates the regularity of the oscillations. Video S6 and at \url{https://youtu.be/1YA5xauI5Y0}.} 
    \label{fig:isolated 40k 16ran 120T}
\end{figure}

Fig.\ref{fig:isolated 40k 16ran 120T} shows the evolution of a bi-partitioned network of 40,000 nodes that are randomly connected to 16 other nodes in the opposite partition. For a narrow band of temperature around \(T_{node} / T_{edge} = 120\) the network state rapidly coheres across the entire scale of the network and periodic oscillatory dynamics emerge.  The period of oscillation depends on temperature (not shown). 

\subsubsection{Externally Biased Networks} \label{sec:externally biased networks}
\begin{figure} [H]
    \centering
    \includegraphics[width=0.3\columnwidth]{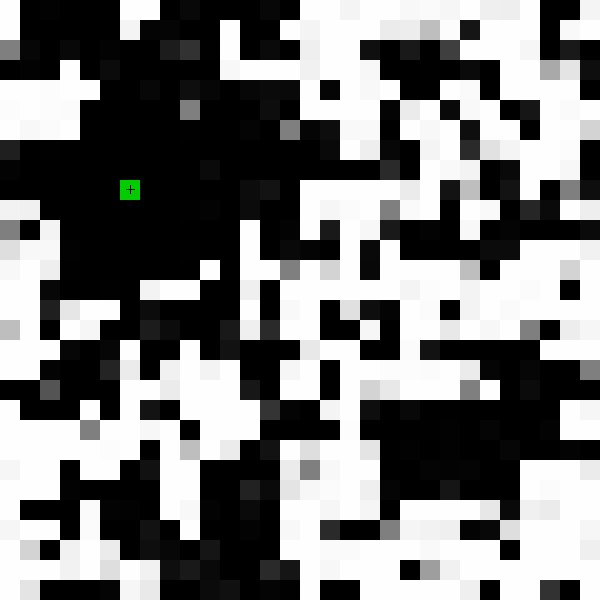}
    \includegraphics[width=0.3\columnwidth]{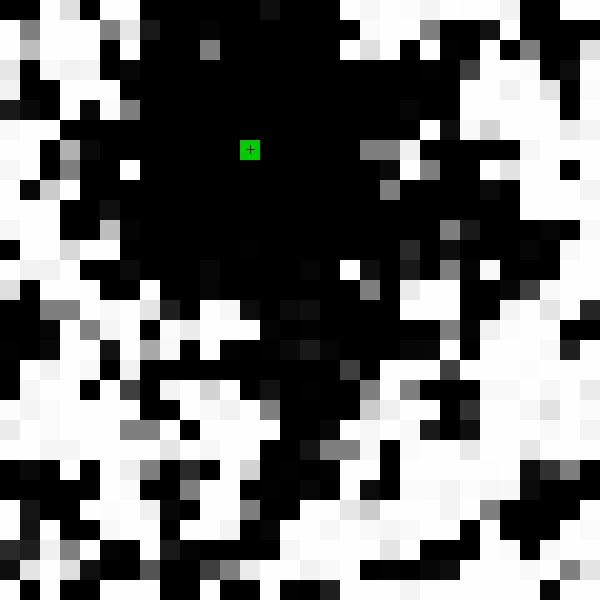}
    \includegraphics[width=0.3\columnwidth]{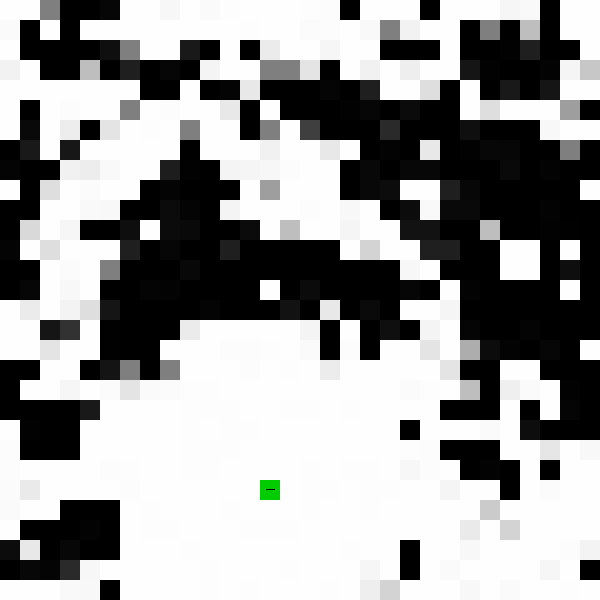}
    \caption{Snapshots from the evolution of a bi-partitioned network of 900 nodes with 4 nearest neighbor connections and a single recurrent connection per node at Tnode / Tedge = 1 with a single externally biased node polarizing the region in its vicinity. The sequence of images is from three different simulations with increasing bias strength (increasing size of fixed edge weights connecting the bias node to its neighbors) from left to right in the images.  Larger bias creates a larger region of polarization.  Domain polarization changes polarity as the biasing node changes sign. Videos S7-S9 and at \url{https://youtu.be/8kiLYNyOMZ8}, \url{https://youtu.be/HD_kJCEqrYA}, \url{https://youtu.be/pZy6S5Huph4}.}
    \label{fig:1 bias 900 node 4nn}
\end{figure}

\begin{figure} [H]
    \centering
    \includegraphics[width=0.38\columnwidth]{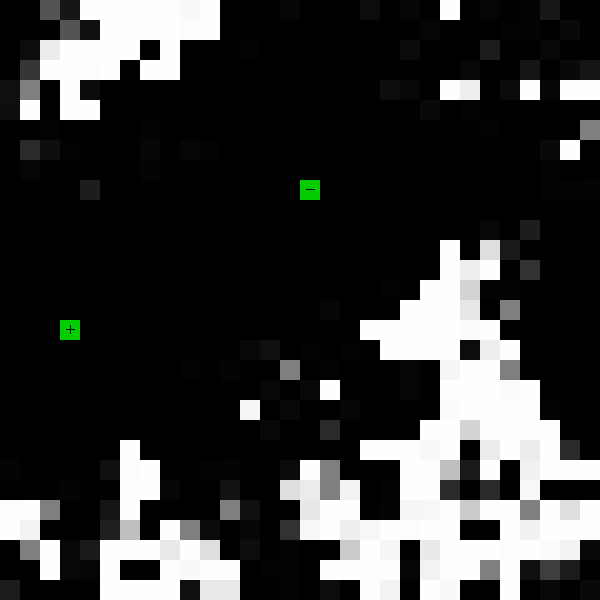}
    \includegraphics[width=0.38\columnwidth]{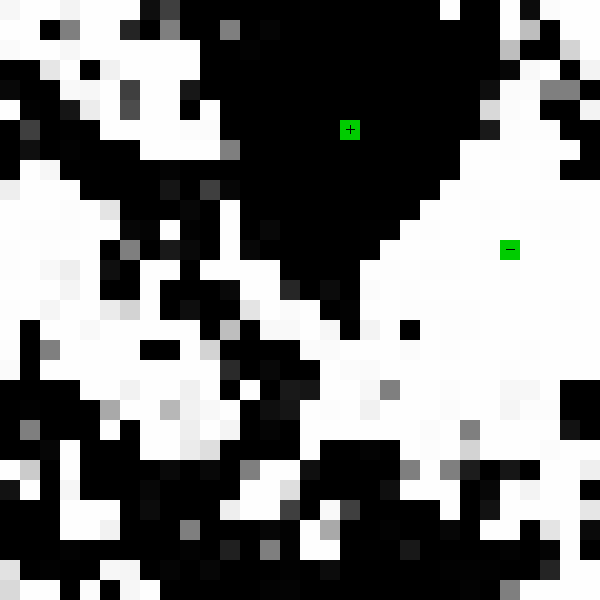} \par
    \includegraphics[width=0.38\columnwidth]{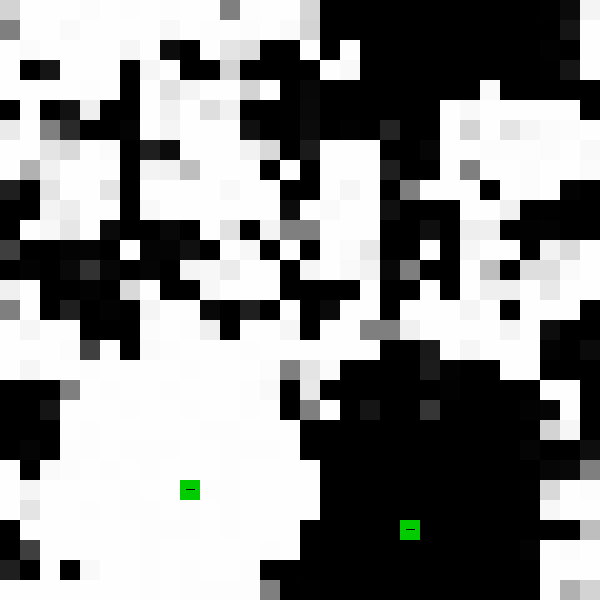}
    \includegraphics[width=0.38\columnwidth]{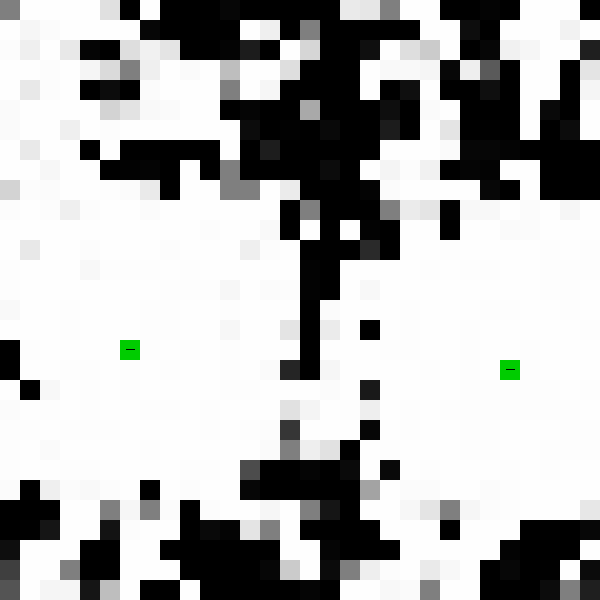}
    \caption{In the same network as in Fig.\ref{fig:1 bias 900 node 4nn}, two biased nodes interact through the network.  If the nodes are of opposite polarity and opposite partition (\textbf{top left}), the network evolves a connection.  If the nodes are of opposite polarity and same partition (\textbf{top right}) or of same polarity and opposite partition (\textbf{bottom left}), the network evolves a domain wall. If the nodes are of the same polarity and same partitions (\textbf{bottom right}), the nodes jointly polarize their surrounding region, but do not grow strong weights between them. Videos S10-S13 and at \url{https://youtu.be/nj-juPln5b0}, \url{https://youtu.be/JdBHyPSUFL0}, \url{https://youtu.be/tVmCSKUA8dQ}, \url{https://youtu.be/2N8zqGX0swM}.}
    \label{fig:2 bias 900 node 4nn}
\end{figure}

\begin{figure} [H]
    \centering
    \includegraphics[width=0.38\columnwidth]{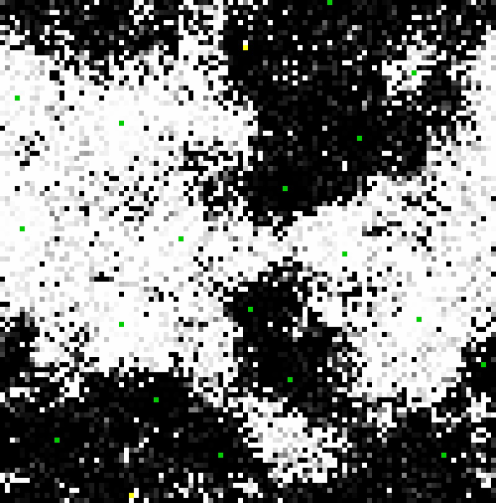}
    \includegraphics[width=0.38\columnwidth]{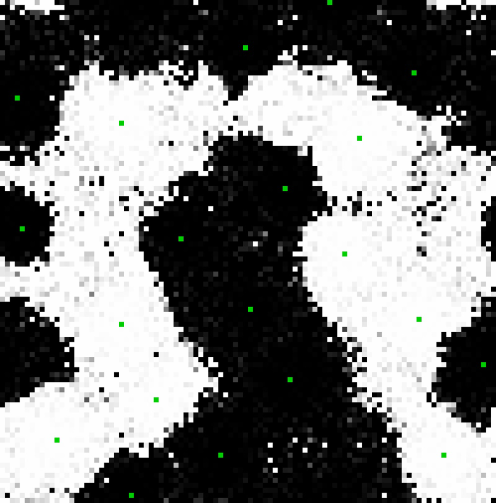} \par
    \includegraphics[width=0.38\columnwidth]{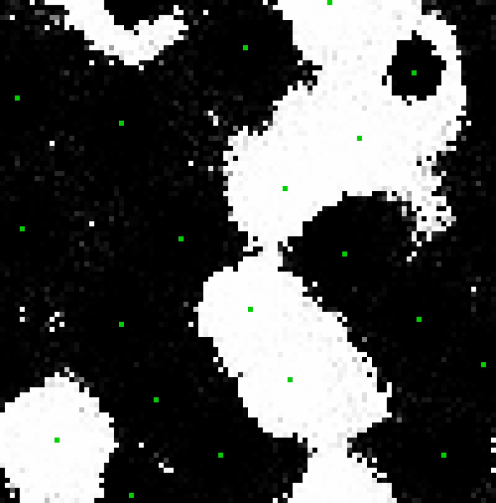}
    \includegraphics[width=0.38\columnwidth]{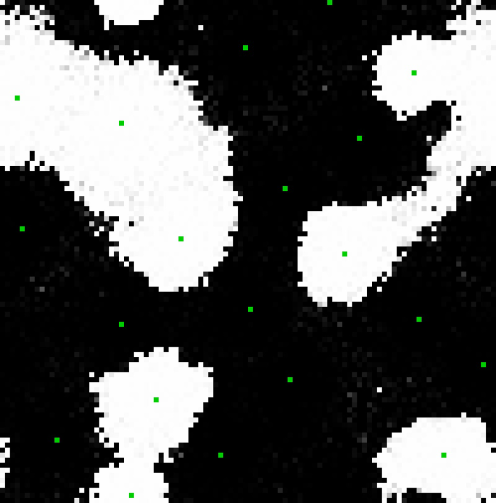}
    \caption{Frames from the evolution of a 10,000 node, bi-partitioned, 16 nearest neighbor network with 2 recurrent connections per node and 10 pairs of bias nodes at \(T_{edge} / T_{node} = 100\).  Each bias pair is composed of complementary nodes (opposite polarity and opposite partition) that change periodically in time, each of the 10 pairs with different periods.  These four images show different configurations of the network as it adapts to changes in its inputs from early to late in the network evolution (left to right and top to bottom).  As the edge weights grow, the domains become sharper and better connected.  As the input nodes change polarity, the network rapidly adapts by creating, destroying and moving domain walls.  In general, the network is challenged to connect and separate nodes into black and white domains according to their polarity and partition (see Fig.\ref{fig:2 bias 900 node 4nn}). Video S14 and at \url{https://youtu.be/xy-eivZ2vJg}.}
    \label{fig:20 bias 10000 node 16nn}
\end{figure}

\begin{figure} [H]
    \centering
    \includegraphics[width=0.4\columnwidth]{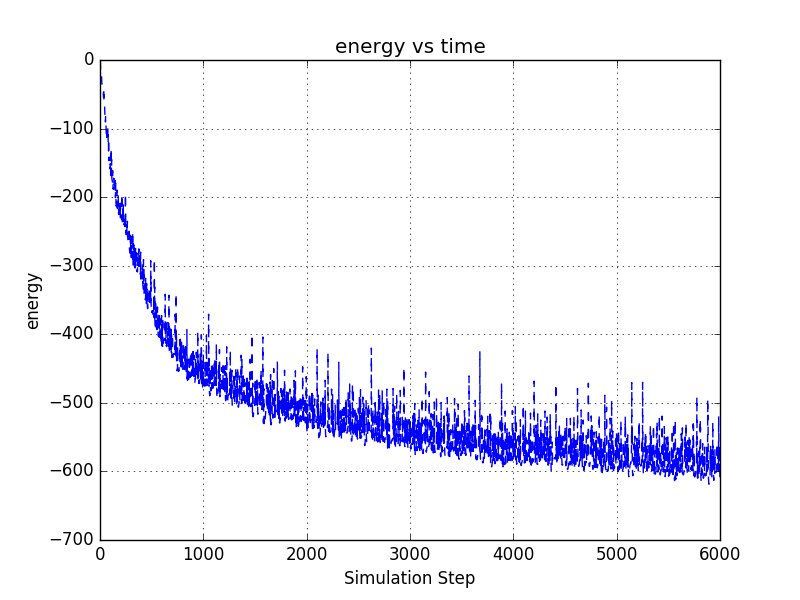}
    \includegraphics[width=0.4\columnwidth]{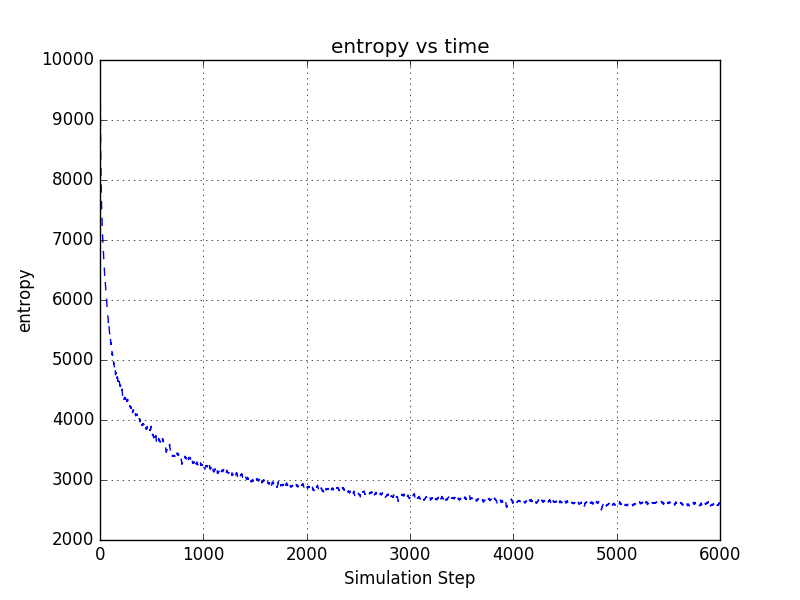} \par
    \includegraphics[width=0.4\columnwidth]{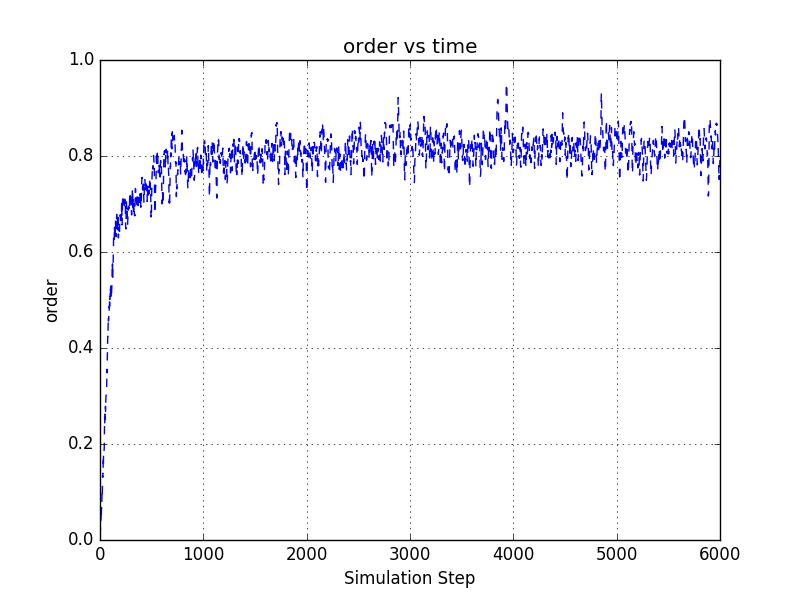}
    \includegraphics[width=0.4\columnwidth]{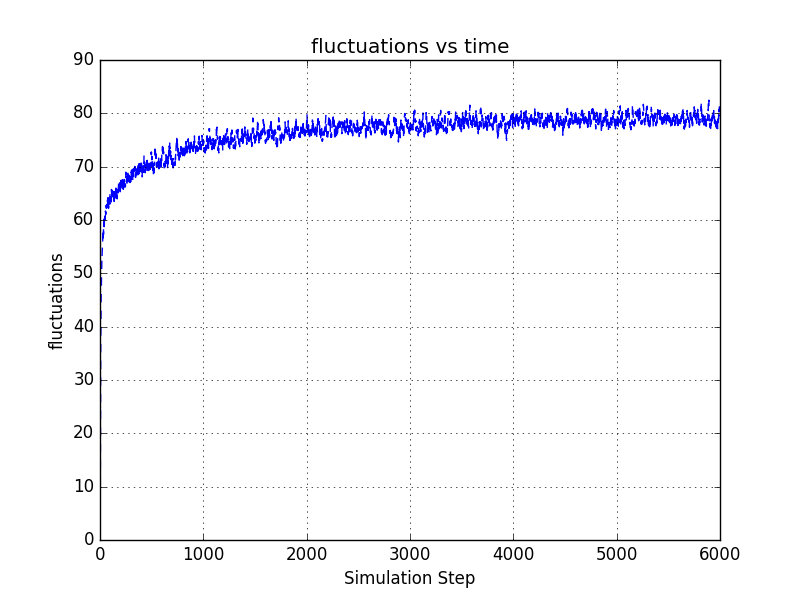}
    \caption{Temporal evolution of selected network averages for the simulation of Fig.\ref{fig:20 bias 10000 node 16nn} – node energy, node entropy, order, and fluctuations vs simulation time.  Energy is the node energy of Eqn.\ref{eqn:node energy 3} averaged over all the nodes.  Entropy is the sum of the node entropies derived from Eqn.\ref{eqn:node state sampling} normalized to a maximum value of 10,000.  Order is average over all edges of (the negative of) the product of the edge’s connected node states.  Fluctuations are the percentage of time that nodes choose a reversible update averaged over all the nodes.  Even as the bias nodes change polarities, the large scale behavior of the network is well behaved.  As compared to the unbiased network example of Figs.\ref{fig:isolated 40k 16nn 100T} and \ref{fig:isolated 40k 16nn 100T statistics}, this network shows lower energy and entropy and a higher degree of order and fluctuation, which is consistent with the application of external bias to the network.}
    \label{fig:20 bias 10000 node 16nn statistics}
\end{figure}

\begin{figure} [H]
    \centering
    \includegraphics[width=0.38\columnwidth]{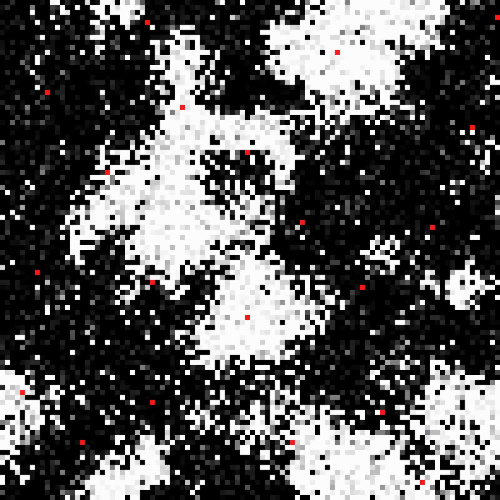}
    \includegraphics[width=0.38\columnwidth]{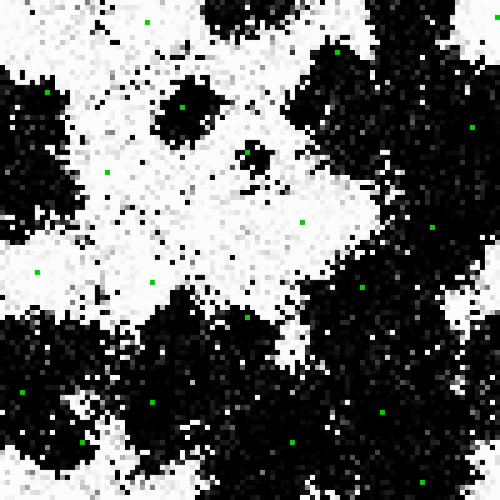} \par
    \includegraphics[width=0.38\columnwidth]{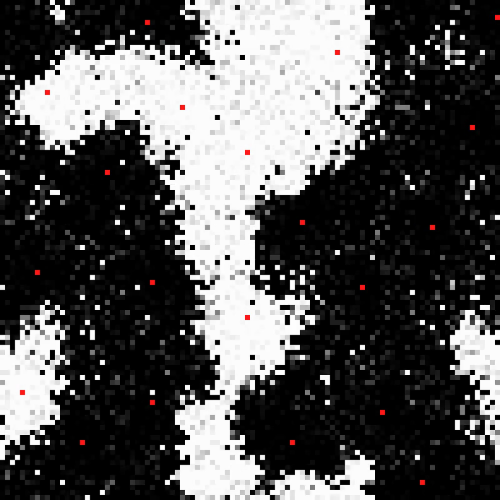}
    \includegraphics[width=0.38\columnwidth]{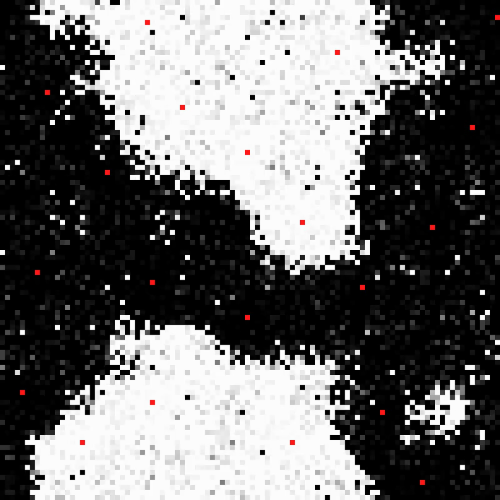}
    \caption{Frames from the evolution of a network like that of Fig.\ref{fig:20 bias 10000 node 16nn} at \(T_{node} / T_{edge} = 120\) and using smaller fixed weights to bias the network. The simulation periodically applies and removes the bias potentials: bias nodes are green when on and red when off. (\textbf{upper left}) The network is not yet exposed to any bias inputs and the dynamics are like those of Fig.\ref{fig:isolated 40k 16nn 120T}. (\textbf{upper right}) Bias has been applied and the network evolves to connect bias nodes as in Fig.\ref{fig:20 bias 10000 node 16nn}. (\textbf{lower left \& right}) With the bias removed the network dynamics continue but are clearly influenced by the modular structure that emerged during the bias phase, seemingly resembling both the unbiased and biased stages of its evolution. Edge weight magnitudes are largely preserved during the unbiased phases of evolution even as they are continuously updated (not shown). Video S15 and at \url{https://youtu.be/ctmWAu09qTE}.}
    \label{fig:20 bias 10000 node 16nn dreaming}
\end{figure}

\begin{figure} [H]
    \centering
    \includegraphics[width=0.4\columnwidth]{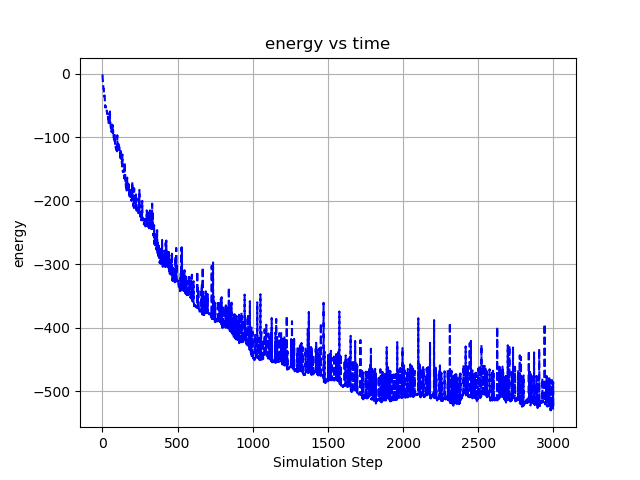}
    \includegraphics[width=0.4\columnwidth]{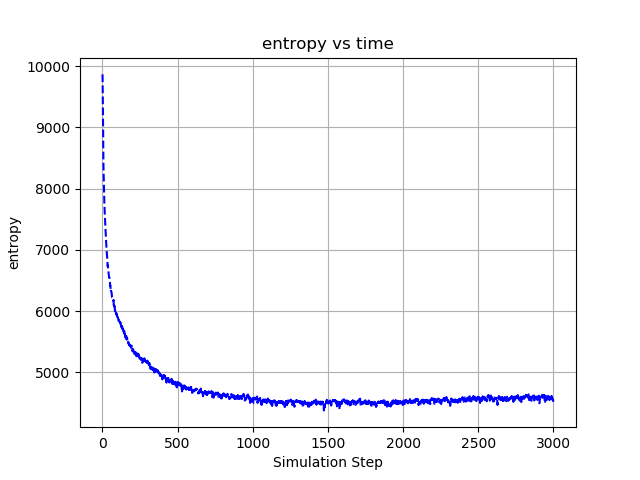} \par
    \includegraphics[width=0.4\columnwidth]{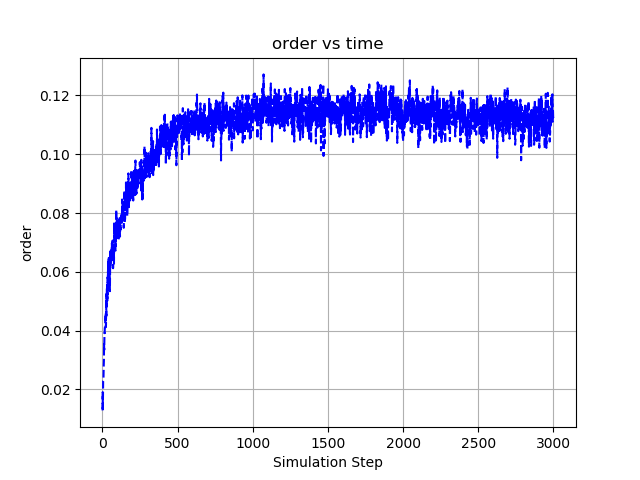}
    \includegraphics[width=0.4\columnwidth]{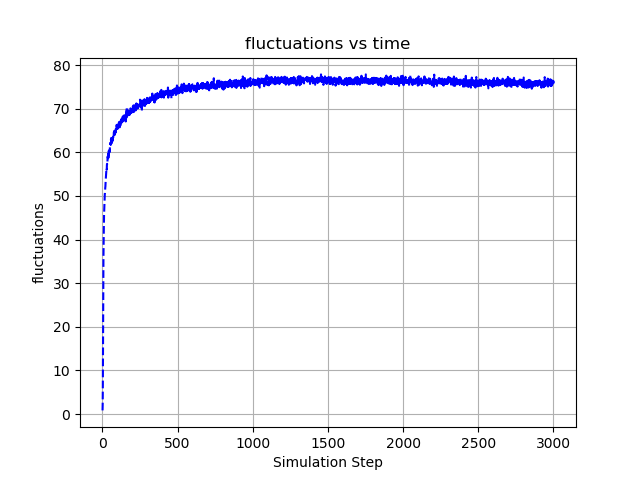}
    \caption{Temporal evolution of selected network averages of a 10,000 node, single partition randomly connected network with 16 connections per node plus 2 recurrent connections per node and 10 pairs of bias nodes at \(T_{edge} / T_{node} = 100\).  Each pair is composed of opposite polarity nodes that change periodically in time, each of the 10 pairs with different periods.  Unlike the networks of Figs.\ref{fig:20 bias 10000 node 16nn} \& \ref{fig:20 bias 10000 node 16nn dreaming}, which are bi-partitioned and locally connected, this network is inherently frustrated and must “carve out” connections amid a vast web of competing interactions. (\textbf{top left}) Energy is the node energy of Eqn.\ref{eqn:node energy 3} averaged over all the nodes. (\textbf{top right}) Entropy is the sum of the nodes entropies from Eqn.\ref{eqn:node state sampling} normalized to a maximum value of 10,000. (\textbf{bottom left}) Order is average over all edges of (the negative of) the product of the edge’s connected node states. (\textbf{bottom left}) Fluctuations are the percentage of time that nodes choose a reversible update averaged over all the nodes.  Although inherently frustrated by it connectivity, as evidenced by it relatively low order, the statistics of the network evolution are similar to those of the nearest-neighbor, bi-partitioned network of Fig.\ref{fig:20 bias 10000 node 16nn statistics}, indicating that the network can effectively evolve connectivity among the bias nodes. Video S16 and at \url{https://youtu.be/PEFAkAcMdVk}.}
    \label{fig:20 bias 10000 node 16ran statistics}
\end{figure}

Figs.\ref{fig:1 bias 900 node 4nn}-\ref{fig:20 bias 10000 node 16ran statistics} are sample results from the simulation of networks biased with external potentials while interacting with a thermal bath.  The panels in Figs.\ref{fig:1 bias 900 node 4nn}, \ref{fig:2 bias 900 node 4nn}, \ref{fig:20 bias 10000 node 16nn}, and \ref{fig:20 bias 10000 node 16nn dreaming} are frames from videos of the network evolution of various bipartite, nearest neighbor networks with periodic boundary conditions. Once again, these examples were chosen because their evolution is easy to visualize. Recursive edges in were employed in these models to avoid local minima in the network nodes as the bias nodes change polarity.

Figs.\ref{fig:1 bias 900 node 4nn} \& \ref{fig:2 bias 900 node 4nn} illustrate the polarization of the network by externally biased nodes and the interaction of externally biased nodes within a network. Fig.\ref{fig:1 bias 900 node 4nn} illustrates the ability of an externally biased node to polarize nodes in its vicinity and communicate with the larger network. Fig.\ref{fig:2 bias 900 node 4nn} illustrates four basic interaction types for two nodes in the network depending on their polarity and their partition placement.  In general the network is able to connect nodes by building strong weights over the long term and to separate nodes by building domain walls in the short term, which is the foundation on which the network can efficiently and rapidly organize itself to transport charge among dynamic external inputs. 

Fig.\ref{fig:20 bias 10000 node 16nn} illustrates the same concepts as Figs.\ref{fig:1 bias 900 node 4nn} \& \ref{fig:2 bias 900 node 4nn} in a much larger network with 10 pairs of externally biased nodes with complementary partitions and complementary potentials that change polarity with different periods.  As it evolves, the network become increasing efficient at segmenting into domains that connect and separate the various bias nodes.  In the video, the propagation of potential through the network is easily visualized as the movement of domains walls in response to changes in the external node polarities.  Selected statistics of the network in Fig.\ref{fig:20 bias 10000 node 16nn} are shown in Fig.\ref{fig:20 bias 10000 node 16nn statistics}, generally showing improvement in the networks performance with time. 

Fig.\ref{fig:20 bias 10000 node 16nn dreaming} replicates the network parameters of Fig.\ref{fig:20 bias 10000 node 16nn} to explore the effect of repeatedly adding and removing external bias potential in a series of "waking" and "sleeping" phases.  As can be seen in the associated video, the sleeping phase retains some of the modular structure (i.e. the coherent regions surrounding each bias node that are similarly polarized) learned during the waking phase.  In this simulation, the external bias weights are small relative to those of the network of Fig.\ref{fig:20 bias 10000 node 16nn} as we found this important in creating complex dynamics in the sleeping phase. 

Fig.\ref{fig:20 bias 10000 node 16ran statistics} shows selected statistics of a single partition randomly connected network otherwise having many features in common with the bi-partitioned nearest neighbor networks of Fig.\ref{fig:20 bias 10000 node 16nn}. By its structure, this network is highly frustrated in its ability to connect nodes and create underlying antiferromagnetic order.  Nonetheless, the statistics of its evolution are qualitatively similar to those of its less frustrated counterpart of Fig.\ref{fig:20 bias 10000 node 16nn statistics}. Video of the network evolution shows evidence of ordering but is much less obvious than in the bi-partitioned, nearest neighbor network examples presented above.

\section{Discussion} \label{sec:discussion}
\subsection{Model Features} \label{sec:model features}
\setlength{\parskip}{0.5em}
\noindent The following paragraphs summarize the main features of the TNN in terms of the physical concepts that motivated its development and the limitations of other artificial neural network paradigms. 

\noindent \textit{Conserved complementary quantities (positive and negative charge) interact but do not cancel.}  Accumulation of charges within the network represent internal potentials (or energetic costs) that can be reduced by transporting these charges through the network and connecting them to external sources and sinks of charge. These ideas are consistent with relations among forces and fluxes in near-equilibrium thermodynamic systems \cite{onsager1931reciprocal}.

\noindent\textit{Fluctuations are inherent in the model formulation.} As compared to most artificial networks, which introduce noise in an ad hoc way, the fluctuations in the TNN are governed by the same relaxation dynamics responsible for charge transport and weight adaptation.  Also the round robin Markov Chain node updates generates fluctuations with correlations that extend through the network.  The fluctuations in the TNN are not "noise" rather, they are thermodynamically consistent, multiscale variations in the network organization.

\noindent\textit{Dissipation of residual charge is coupled to fluctuations.} Mismatch of input charges on a node creates residual charge that must be dissipated as the network equilibrates.  The edge weight updates (\ref{eqn:residual charge split 3}) strive to eliminate this mismatch, but thermal noise in the updates means that the match is always imperfect - even in networks driven by constant external inputs evolved to a low energy steady state.  Hence, dissipation in the network is inextricably linked to fluctuations and charge transport "resistance" is emergent and unavoidable.  Fluctuation-dissipation effects are well known from equilibrium statistical physics \cite{callen1951irreversibility} and recently extended to non-equilibrium systems \cite{jarzynski1997nonequilibrium} \cite{crooks1999entropy}.

\noindent \textit{Adaptation is coupled to the dissipation of conserved quantities.} While updating weights to eliminate errors in an objective function is the foundation of most neural network models, in the TNN model it also has a physical interpretation.  Namely, the dissipation of conserved quantities within an open physical system (e.g. the residual node charges for the TNN), when coupled to the system features responsible for their creation (e.g. the edge weights), can adapt the system to reduce dissipation under similar future conditions.  If the environment in which this system is embedded has certain stable features in the potentials presented to the system, then through its interaction with that environment, the system may come to represent and predict those features and to thereby minimize internal dissipation \cite{still2012thermodynamics}.  The intuition here is that the dissipation of a conserved physical quantity requires a physical structure to transport it out of the system.  In the TNN model, this supposition is that the edge weight, which creates the charge imbalance, mediates the transport of the residual charge to the reservoir, and in the process is adapted by it.  As an example from everyday life, consider a housing construction site in which certain raw materials (the conserved quantities) are cut as the house is built and residual scraps of material are produced that cannot be used.  Those scraps, which must be transported away from the construction site (the dissipation), can be used to inform the acquisition of materials in future (the adaptation) and to improve the efficiency of the construction process up to the point that variances in materials and construction permit (the fluctuation). This process can become highly predictable and efficient if the same house is constructed many times, materials suppliers are reliable, and labor is consistent (a stable environment).

\noindent\textit{Edge states adapt with respect to the current state of their connected nodes without destroying state information associated with other node states.} The kinetic factors in the model adapt edge states selectively depending on the node state (Eqns.\ref{eqn:node energy 3}, \ref{eqn:residual charge} \& \ref{eqn:residual charge split 3}).  We presume that this techniques allows edge state updates associated with somewhat different configurations of the collective node states to reinforce to the degree that they are similar but also not cancel to the degree that they are different, thereby enforcing commonalities (generalization) while preserving differences (specialization).  Referring again to the example of constructing a house, building a wall and building a floor may share similar tools and fasteners, but use different types of lumber.  In adapting for these tasks we would like to generalize the adaptation for tools and fasteners while specializing the adaptation for the different types of lumber. 

\noindent\textit{Rapid, global relaxation of the nodes states and slower, local adaptation of the edge states results in the evolution of a multiscale, complex system.}  Successful adaptation requires that the network achieve equilibrium as quickly as its inputs change and that it refine its connectivity with time to improve this ability to equilibrate in the future, which are implemented through the reversible and irreversible node update decisions, respectively.  In the house building analogy, a reversible update might involve the distribution of the materials for the day’s work while an irreversible update might involve the many individual activities using those materials to construct the house.  Building the house efficiently requires both that the materials are well distributed and that future distributions are refined according to the waste produced in the construction process.  A variety of natural, networked systems and models of such systems involve the idea of adaptation at different scales. Comparison of various systems related to physics, materials, ecology, biology and cognition indicate that “dual phase evolution” may explain common observations of modularity, network statistics and criticality \cite{paperin2010dual} in these diverse domains. We speculate that this idea my also be responsible for the abilities of brains to rapidly orient to new environments and to learn rapidly learn from unfamiliar experiences and place them in their correct context (so called "one shot" learning).

\noindent\textit{The TNN employs concepts from equilibrium statistical physics to evolve the dynamic organization of a model system that may be in or out of equilibrium.} Both the reversible and irreversible updates to the TNN (Eqns.\ref{eqn:node state sampling} \& \ref{eqn:residual charge split 3}) employ a Boltzmann distribution to sample node and edge states.  Further, these distributions are not employed to compute statistics, but to drive the dynamics of network evolution.  While the overall validity of such an approach as a model of natural phenomena is not clear, all state decisions in the modal are local to the nodes and, as such, local equilibria may be the correct context for updating a distributed, interacting network like the TNN. 

\noindent\textit{Large scale stochastic dynamics can emerge through local interactions.}  The videos referenced in Figs.\ref{fig:isolated 40k 16nn 100T}, \ref{fig:20 bias 10000 node 16nn} \& \ref{fig:20 bias 10000 node 16nn dreaming}, for example, clearly show large scale stochastic dynamics.  The node-to-edge temperature ratio, bias strength, and network connectivity also play crucial roles in the large scale dynamics of the networks.  Although not emphasized in the work presented here, sharp transitions from order to disorder as temperature changes, phase transitions, are also possible.

\noindent\textit{Networks evolve causal dynamics related to the spatial and temporal structure of potentials in their environment.} As external potentials change in time, the network responds with corresponding changes in it spatial and temporal structure, which can be characterized as the potentials \textit{causing} change within the network via the transport of charge through it. The round robin Markov chain technique insures temporal consistency because every node update is preceded by an update of all the other nodes in the network.  The irreversible updates, occurring when the nodes are in a local equilibrium, insure that the organization learned by the network is that which is consistent with the second law of thermodynamics and the causative "arrow of time". 

\noindent\textit{The TNN avoids many computational challenges found in other neural network models.}  Nodes can be connected as networks of any type without creating dynamic instabilities.  Node and edge updates are continuous and online without forward and backward passes; there is no separation of “learning” and “inference”.  There are few ad hoc meta-parameters: for example, there are no learning rates.  There are no gradients that need to be computed or communicated across layers.

\noindent\textit{The TNN unifies concepts of conservation, potentiation, fluctuation, dissipation, adaptation, equilibration and causation under a common physical model to illustrate a multiscale, self-organizing, complex, adaptive system.}  The model self-organizes with and without external inputs.  Externally applied potentials propagate through the network by polarizing connected nodes.  Self-organization is strongly modulated by network effects, the relative temperatures of the nodes and edges, and the strength of the external applied potentials.

\subsection{Limitations and Speculation on Future Opportunities} \label{sec:model limitations}

The most challenging part of the implementation of the model is the search of the network space to find a representative, low-energy state in the Markov Chain round robin.  As is typical, the search for a global optimum is frustrated by local minima and there is no all-purpose algorithm to address this problem \cite{wolpert1997no}.  In the implementation described here, this is typically recognized as a domain that fails to change state as the external potentials transition (which is partially mitigated by the recursive edges in the examples of Sec.\ref{sec:externally biased networks}).  In the videos associated with Sec.\ref{sec:externally biased networks}, these failures are sometimes visible directly in the networks and are also explicity illustrated by the yellow color of the bias nodes when any of their connected nodes have the wrong polarity.  There is little doubt that the methods used here might be improved to address this challenge, but more comprehensive searches also face a combinatorial explosion of potential evaluations. 

It is interesting to consider the source of this challenge in the context of the thermodynamic concepts that motivated the TNN.  Every computing model is composed of a sequence of variable assignments.  The ability to make these assignments requires that the variables of the model be independent at the time of assignment.  For example, if we wish to perform the assignment \(a \gets b \cdot c \) then \(b\) and \(c\) must exist and be independent of \(a\) at the time of the assignment.  In the model implementation described here, this limitation is recognized in the constant node-at-a-time round-robin search for a low energy configuration of the node states as in Eqn.\ref{eqn:node state sampling} and in the approximations leading to the edge weight updates Eqns.\ref{eqn:residual charge split 2}, \ref{eqn:residual charge split 3} \& \ref{eqn:weight reduction}.  More generally, the challenge of creating the TNN can be seen as taking the relatively simple statement of Eqn.\ref{eqn:network equilibrium} and the concepts of \ref{sec:model concepts} and translating them into a sequence of variable assignments that effectively addresses the challenges of capturing the interdependencies of the state variables.  The computational techniques described above are both critical to the implementation of the TNN model as well as the illustration of its greatest difficulty.

While we can claim some success in our efforts to address the challenge just described and suppose even that there might be useful implementation of the TNN, there are certain ironies implicit in simulating complex thermodynamic systems on deterministic computing hardware.  We must, for example, calculate probability distributions and generate pseudo-random numbers to sample fluctuations using computing hardware that, at great expense, is engineered, manufactured and operated to prevent fluctuations.  We must, for example, at great expense, search for a representative sample of an equilibrium distribution, while every natural system does this at essentially zero cost.  So, perhaps the most promising future implementations of models such as the one presented here would involve hardware in which the device electronics inherently perform the thermodynamic relaxation that drives the evolution of the network.  For example, nodes might be constructed of multistate devices that are marginally stable at their operating temperature and that can be biased to favor transition to a particular state by the charge received from their inputs.  Also, edges might be constructed of semi-stable, hysteretic resistive components (“memristors” or “memcapacitors”) that change impedance depending on the history of the current passing through them \cite{wang2017memristors}.  Such systems would have orders of magnitude higher energy efficiency, scalability and perhaps offer much more complex functionality than the computational model described here.  These future systems, in combination with conventional computing elements, might create the foundations for a “thermodynamic computer” \cite{hylton2019thermodynamic-computing} that can both evolve “from below” according to the basic thermodynamics of its components and be constrained “from above” by human specified code.  In such systems “thermodynamic evolution” might be an omnipresent capacity driving its self-organization toward a high-level, human specified goal.

\section{Materials and Methods} \label{sec:materials and methods}

These results are the product of simulation on a laptop computer.  The implementation stresses flexibility in the exploration of ideas and has not been refined for efficiency of execution or speed. There is a very high degree of parallelism in the model that very likely could be effectively executed in modern architecture like multi-core CPUs, GPUs, and emerging asynchronous and neuromorphic computing systems \cite{debole2019truenorth} (DeBole, et al., 2019). The code is available at \url{https://github.com/toddhylton/Thermodynamic-Neural-Network---Public}. 

\section{Conclusions} \label{sec:conclusions}
We have described a neural network model comprising a collection of the nodes and edges that that organizes according to basic principles of physics and thermodynamics.  Charge conservation laws and the hypothesis that nodes should evolve to transport charge effectively results in networks of nodes that organize to maximize charge transport efficiency.  Node and edge state updates derive from relaxation of the network according to Boltzmann statistics.  Node states relax globally and reversibly in concert with edge states that relax locally and irreversibly, resulting in a multiscale self-organizing, complex system with dynamics that are sensitive to network structure and temperature.  Externally applied potentials diffuse into the network, establishing strong connections to complementary potentials and creating domain walls to separate competing potentials.  The model integrates ideas of conservation, potentiation, fluctuation, dissipation, adaptation, equilibration and causation to illustrate the thermodynamic evolution of organization.

\vspace{6pt} 

\supplementary{The following are available online at \linksupplementary{s1}}
\begin{itemize}
\item Video S1: {Fig.\ref{fig:isolated w/ antiferro domains} network evolution}\label{video:S1}
\item Video S2: {Fig.\ref{fig:isolated w/ ferro domains} network evolution}\label{video:S2}
\item Video S3: {Fig.\ref{fig:isolated 40k 16nn 80T} network evolution}\label{video:S3}
\item Video S4: {Fig.\ref{fig:isolated 40k 16nn 100T} network evolution}\label{video:S4}
\item Video S5: {Fig.\ref{fig:isolated 40k 16nn 120T} network evolution}\label{video:S5}
\item Video S6: {Fig.\ref{fig:isolated 40k 16ran 120T} network evolution}\label{video:S6}
\item Video S7: {Fig.\ref{fig:1 bias 900 node 4nn} network evolution w/ small bias weights}\label{video:S7}
\item Video S8: {Fig.\ref{fig:1 bias 900 node 4nn} network evolution w/ medium bias weights}\label{video:S8}
\item Video S9: {Fig.\ref{fig:1 bias 900 node 4nn} network evolution w/ large bias weights}\label{video:S9}
\item Video S10: {Fig.\ref{fig:2 bias 900 node 4nn} network evolution w/ opposite polarity \& opposite partition bias nodes}\label{video:S10}
\item Video S11: {Fig.\ref{fig:2 bias 900 node 4nn} network evolution w/ opposite polarity \& same partition bias nodes}\label{video:S11}
\item Video S12: {Fig.\ref{fig:2 bias 900 node 4nn} network evolution w/ same polarity \& opposite partition bias nodes}\label{video:S12}
\item Video S13: {Fig.\ref{fig:2 bias 900 node 4nn} network evolution w/ same polarity \& same partition bias nodes}\label{video:S13}
\item Video S14: {Fig.\ref{fig:20 bias 10000 node 16nn} network evolution}\label{video:S14}
\item Video S15: {Fig.\ref{fig:20 bias 10000 node 16nn dreaming} network evolution}\label{video:S15}
\end{itemize}


\funding{This research received no external funding}

\acknowledgments{The inspiration for this work started roughly 12 years ago, shortly before I began working at DARPA.  Since that time many people have influenced my thinking in the domains of thermodynamics, learning, neural networks, complex systems, computation, cognitive science, neuroscience and related domains that were key to the development of the TNN.  I would like to acknowledge these people in particular for their insights and support – Alex Nugent, Yan Yufik, Yaneer Bar Yam, Robert Fry, Ben Mann, Dharmendra Modha, Narayan Srinivasa, Stan Williams, Jennifer Klamo, Filip Piekniewski, Patryk Laurent, Csaba Petre, Micah Richert, Dimitri Fisher, Tom Conte, Michael Hazoglou, Natesh Ganesh, and Ken Kreutz-Delgado.}

\conflictsofinterest{The author declares no conflict of interest.} 
\reftitle{References}
\externalbibliography{yes}
\bibliography{TNN}



\end{document}